\documentclass[11pt, a4paper]{article}
\usepackage{jheparxiv}
\usepackage[utf8]{inputenc}
\usepackage{amsmath}
\usepackage{amsfonts}
\usepackage{amssymb}
\usepackage{latexsym}
\usepackage{mathrsfs}
\usepackage{braket}		
\usepackage{graphicx}
\usepackage{color}
\usepackage{xcolor}
\usepackage{slashed}
\usepackage{twistor}
\usepackage[all]{xy}
\usepackage{tikz-cd}

\newcommand{\sa}{\mathsf{a}}

\newcommand{\rg}{\mathrm{g}}

\newcommand{\sT}{\mathsf{T}}

\newcommand{\ba}{\mathbf{a}}

\renewcommand{\d}{\mathrm{d}}

\newcommand{\qed}{\hfill \ensuremath{\Box}}

\newcommand{\eps}{\epsilon}
\newcommand{\Dbar}{\bar{D}}
\newcommand{\g}{\mathfrak{g}}
\newcommand{\msf}[1]{\mathsf{#1}}
\newcommand{\F}{\mathcal{F}}

\newcommand{\sH}{\mathsf{H}}

\newcommand{\scG}{\mathscr{G}}

\title{Gluon scattering on self-dual radiative gauge fields}

\author[a]{Tim Adamo,}
\author[b]{Lionel Mason}
\author[b]{\& Atul Sharma}

\affiliation[a]{School of Mathematics and Maxwell Institute for Mathematical Sciences \\
        University of Edinburgh, EH9 3FD, United Kingdom}
        
\affiliation[b]{The Mathematical Institute \\
		University of Oxford, Woodstock Road, OX2 6GG, United Kingdom}

\emailAdd{t.adamo@ed.ac.uk}
\emailAdd{[lmason,atul.sharma]@maths.ox.ac.uk}

\abstract{We present all-multiplicity formulae, derived from first principles in the MHV sector and motivated by twistor string theory for general helicities, for the tree-level S-matrix of gluon scattering on self-dual radiative backgrounds. These backgrounds are chiral, asymptotically flat gauge fields characterised by their free radiative data, and their underlying integrability is captured by twistor theory. Tree-level gluon scattering scattering amplitudes are expressed as integrals over the moduli space of holomorphic maps from the Riemann sphere to twistor space, with the degree of the map related to the helicity configuration of the external gluons. In the MHV sector, our formula is derived from the Yang-Mills action; for general helicities the formulae are obtained using a background-coupled twistor string theory and pass several consistency tests. Unlike amplitudes on a trivial vacuum, there are residual integrals due to the functional freedom in the self-dual background, but for scattering of momentum eigenstates we are able to do many of these explicitly and even more is possible in the special case of plane wave backgrounds. In general, the number of these integrals is always less than expected from standard perturbation theory, but matches the number associated with space-time MHV rules in a self-dual background field, which we develop for self-dual plane waves.}

\begin{document}

\maketitle
  
\section{Introduction}

There are many physical scenarios which are well approximated by perturbative quantum field theory (QFT) in \emph{strong background fields}: fixed, non-trivial solutions to the classical equations of motion which are treated non-perturbatively~\cite{Furry:1951zz,DeWitt:1967ub,tHooft:1975uxh,Abbott:1981ke}. In the context of gauge theories, strong Maxwell or Yang-Mills background fields play an important role in the study of laser physics (cf., \cite{DiPiazza:2011tq,King:2015tba,Seipt:2017ckc}), the colour-glass condensate effective theory of heavy ion collisions (cf., \cite{Iancu:2002xk,Gelis:2010nm}) and rapidity evolution of Wilson lines modeling high-energy hadron collisions (cf., \cite{Balitsky:1995ub,Balitsky:2004rr,Caron-Huot:2013fea}). However, the computation of physical observables such as scattering amplitudes in these scenarios is notoriously difficult due to the complicated nature of perturbative Feynman rules defined by the background field Lagrangian. Consequently, there are no tree-level calculations beyond four external states for QFTs in even the simplest strong gauge field backgrounds\footnote{Examples of such 4-point calculations include trident pair production in strong field QED (cf., \cite{Hu:2010ye,Ilderton:2010wr,King:2013osa,Dinu:2017uoj,Mackenroth:2018smh}) or the 4-gluon tree amplitude in a Yang-Mills plane wave background~\cite{Adamo:2018mpq}.}.

By contrast, recent decades have seen the development of myriad all-multiplicity formulae for the scattering amplitudes of gauge theory in trivial backgrounds at tree-level and beyond. Crucially, these all-multiplicity results are not derived from standard space-time perturbation theory. This raises the question: is it possible to derive all-multiplicity results in strong field Yang-Mills theory by exploiting alternatives to background perturbation theory on space-time?

In four space-time dimensions, one powerful example of such an alternative is \emph{twistor theory}; an algebro-geometric formalism that (very roughly) encodes physical fields in geometric data on a complex projective variety called twistor space. Twistor theory enables gluon scattering to be formulated as a perturbative expansion around the integrable self-dual sector of gauge theory. This can be operationalized in terms of a `twistor string' governing holomorphic maps from the Riemann sphere to twistor space, with the degree of the map corresponding to the helicity configuration of the external gluons in a scattering process (i.e., how `far away' from self-duality the configuration is)~\cite{Witten:2003nn,Berkovits:2004hg}. As such, twistor theory underpins one of the most well-known (and earliest) examples of an all-multiplicity formula for scattering amplitudes: the Roiban-Spradlin-Volovich-Witten (RSVW) formula for the full tree-level S-matrix of Yang-Mills theory (in a trivial background)~\cite{Roiban:2004yf}.
 
\medskip

\emph{A priori}, it might seem unlikely that a mathematical formalism like twistor theory could say anything about strong field gauge theory, a topic usually studied with the brute-force methods of background-dressed Feynman rules on space-time. However, the underlying chirality of twistor theory means that it can be used to essentially trivialize the self-dual sector of Yang-Mills theory, which means that it should be applicable to scattering on self-dual backgrounds in much the same way as a trivial background.

In this paper, we present formulae for all tree-level gluon scattering amplitudes in any self-dual (SD) radiative Yang-Mills background. These backgrounds are chiral gauge fields which are characterized by their free radiative data and are asymptotically flat (or `almost everywhere' asymptotically flat); as such, they admit a well-defined S-matrix. An important special case of such SD radiative backgrounds are the self-dual plane waves, which can be viewed as a coherent superposition of positive helicity gluons valued in a Cartan subalgebra of the gauge group. 

Despite having unconstrained functional freedom, these SD radiative backgrounds -- and gluon perturbations propagating on them -- admit an elegant description in terms of twistor theory. For the maximal helicity violating (MHV) sector (i.e., with two negative helicity external gluons and the remainder positive helicity), this allows us to derive -- from first principles -- momentum space expressions for the corresponding tree amplitudes at arbitrary multiplicity; these include as a special case the formula of~\cite{Adamo:2020syc} for SD plane wave backgrounds. We also provide formulae for all-multiplicity tree-level gluon scattering in \emph{any} helicity configuration on a SD radiative background which are motivated by twistor string theory. While the formulae for these N$^k$MHV amplitudes are conjectural for $k\geq2$, they are easily seen to pass several fundamental consistency tests.    

Although the non-trivial background obviously introduces new features, our formulae share a surprising amount of structure with their well-known antecedents on a trivial background, and manifest remarkable simplicity in comparison to the expectations of space-time perturbation theory. For instance, in the special case where the SD radiative background is Cartan-valued, the MHV colour-ordered partial amplitude is given by:
\begin{equation}
 \frac{\la r\,s\ra^4}{\la1\,2\ra\,\la2\,3\ra\,\cdots\la n-1\;n\ra\,\la n\,1\ra}\int\d^4x\;\exp\left[\sum_{i=1}^n \left(\im\,k_i\cdot x + e_i\,g(x,\kappa_i)\right)\right]\,,
\end{equation}
where the external gluons are parametrized by their on-shell asymptotic 4-momenta $k_{i}^{\alpha\dot\alpha}=\kappa_i^{\alpha}\,\tilde{\kappa}_{i}^{\dot\alpha}$ and charges $e_i$ with respect to the background, and gluons $r$, $s$ are negative helicity with all others positive helicity. In this expression, the ratio of angle brackets (constructed from the un-dotted momentum spinors, $\la i\,j\ra=\kappa_{i}^{\alpha}\,\kappa_{j\,\alpha}$) is familiar from the Parke-Taylor formula~\cite{Parke:1986gb} for MHV gluon scattering in a trivial background. But 4-dimensional momentum conservation is absent (as Poincar\'e invariance is broken by the background), replaced by a single space-time integral against an exponential factor, where the Cartan-valued function $g(x,\kappa)$ is determined by the SD background.

For a N$^{d-1}$MHV amplitude on a generic SD radiative background, our central conjecture is the formula (presented with $\cN=4$ supersymmetry for simplicity)
\begin{equation}
\cA_{n,d}=\int \frac{\d^{4|4(d+1)}U}{\mathrm{vol}\,\mathrm{GL}(2,\C)}\,\tr\left(\prod_{i=1}^{n}\frac{\sH_i^{-1}\,a_{i}\, \sH_i \,\D\sigma_i}{(i\,i+1)}\right) \,,
\end{equation}
where the integral is over the moduli of a degree $d$ holomorphic map from the Riemann sphere to twistor space, the $a_i$ are twistor wavefunctions for the gluon multiplets, and $\sH_i$ are certain holomorphic frames which encode the information about the background. In a trivial background, $\sH_{i}=\mathrm{id}$, and this expression immediately reduces to the RSVW formula. In general, there are $4d$ residual (bosonic) integrals in this formula, but when the external gluons are represented by momentum eigenstates, $2(d+1)$ integrations can be done explicitly against delta functions (see \eqref{Cart3}), yielding a formula supported on $2(n-d-1)$ of the refined/polarized scattering equations known from ambitwistor strings in trivial backgrounds~\cite{Geyer:2014fka}. In the special case of a self-dual plane wave background, the number of residual integrals can be reduced to $2d-1$ (see \eqref{YMamp-pw-fin}) due to enhanced symmetry. 

Such residual integrals are a generic feature of amplitudes in strong fields, as the background generically breaks Poincar\'e invariance and introduces functional degrees of freedom. However, even in the most general case, the $4d$ residual moduli integrals of our formula are less than the number of integrations expected from space-time perturbation theory in every helicity configuration for $n>3$. This number of integrations is precisely what would be generated for a space-time version of the MHV formalism~\cite{Cachazo:2004kj} in a non-trivial background, and in section~\ref{discussion} we give the full off-shell propagator in MHV axial gauge in a self-dual plane wave background. Combined with our MHV amplitudes, this yields an MHV formalism on such backgrounds.

\medskip

The paper is organized as follows. Section~\ref{SECT:Twistor} begins with an overview of SD radiative gauge fields and their twistor theory. Section~\ref{SECT:MHVgauge} derives the MHV gluon amplitude on a SD radiative background from a first-principles using a space-time generating functional. The resulting formula is shown to pass basic consistency tests arising from the weak-field background limit. In section~\ref{SECT:NMHV}, we present a formula for the complete gluon tree-level S-matrix on a SD radiative gauge field background, motivated by twistor string theory, which also passes these basic consistency tests. 

Throughout, the special cases of Cartan-valued backgrounds and self-dual plane waves are used as illustrative and explicit examples. Appendix~\ref{App:YM} includes Feynman diagram computations in space-time perturbation theory which match our formulae at 3- and 4-points, while appendix~\ref{App:Imp} evaluates the formula for the tree-level S-matrix in the special case of an impulsive self-dual plane wave background, where interesting simplifications emerge. Lastly, appendix~\ref{propapp} develops MHV rules on SD plane wave backgrounds which provide an explanation for the number of residual integrals in our formulae.


\section{Twistor theory of self-dual radiative gauge fields}
\label{SECT:Twistor}

A (non-linear) gauge field in Minkowski space is defined to be purely \emph{radiative} if it is source-free, asymptotically flat at both past and future conformal null infinity $\scri^-\cup \scri^+=\scri$ and completely determined by the free characteristic data at either $\scri^-$ or $\scri^+$. In four dimensions, this radiative data is encoded in two functions on $\scri$ valued in the adjoint representation of the gauge group. The natural observable of Yang-Mills theory in such a configuration is the S-matrix, defined perturbatively on the radiative background by the scattering map of data from $\scri^-$ to that at $\scri^+$. The reader uninterested in precise definitions of radiative self-dual gauge fields can skip section~\ref{sec:scri}, proceeding to section~\ref{YMTwTh} which developes the twistor theory of self-dual radiative backgrounds.

Complexifying the data and background field, one of the free radiative functions on $\scri$ can be set to zero to obtain a self-dual (SD) solution of the Yang-Mills equations. Such solutions can only be real in split or Euclidean signature, but we will not impose such reality conditions; complex circularly polarized solutions have meaning in the quantum theory. In four dimensions, the SD sector is famously integrable: SD solutions can be constructed from complex analytic data with twistor theory~\cite{Mason:1991rf}. For SD radiative gauge fields, there is a twistor construction which manifests their integrability in terms of the characteristic data at $\scri$~\cite{Sparling:1990}. In this section, we review this twistorial description of SD radiative gauge fields and demonstrate how gluon perturbations on these backgrounds are encoded in their twistor theory. Throughout, we highlight the concrete example of self-dual plane waves (SDPWs); while not strictly asymptotically flat, as the fields do not fall off fast enough in the null symmetry direction, they are sufficiently `almost asymptotically flat' for scattering to be well-defined~\cite{Schwinger:1951nm,Adamo:2017nia}.


\subsection{Radiative gauge fields and self-duality}
\label{sec:scri}

\paragraph{Spinor conventions and Null infinity, $\scri$:} The 2-spinor formalism (cf., \cite{Penrose:1984uia}) is introduced by writing
\be\label{coord}
x^{\alpha\dot\alpha}  := \frac{1}{\sqrt{2}}\begin{pmatrix}t+x^3&&x^1-\im\,x^2\\x^1+\im\,x^2&&t-x^3\end{pmatrix},
\ee
where $x^a=(t,x^1,x^2,x^3)$ are the standard Cartesian coordinates on Minkowski space $\R^{1,3}$ or its complexification $\M=\C^4$. Here, $\alpha,\dot\alpha$ are two-component spinor indices associated to the spinor representations $\mathfrak{so}(1,3)\cong\mathfrak{sl}(2)\oplus\widetilde{\mathfrak{sl}(2)}$. The spinor components of a vector $v^a$ are also defined via \eqref{coord}. The metric on complex Minkowski space, $\M$, is given by 
\be\label{eta}
\d s^2 = \d x^{\alpha\dot\alpha}\,\d x_{\alpha\dot\alpha},
\ee
where spinor indices are raised and lowered using the two-dimensional Levi-Civita symbols $\epsilon_{\alpha\beta}=\epsilon_{[\alpha\beta]}$, $\epsilon_{\dot\alpha\dot\beta}=\epsilon_{[\dot\alpha\dot\beta]}$, $\epsilon_{01}=1$, etc.. In the spinor helicity formalism, contractions of spinors of each chirality are denoted by $\la a\,b\ra := \eps^{\alpha\beta}a_\beta b_\alpha$ and $[a\,b] := \eps^{\dot\alpha\dot\beta}a_{\dot\beta}b_{\dot\alpha}$. 

In four-dimensions, a 2-form $F$ decomposes into self-dual (SD) and anti-self-dual (ASD) parts as: 
\be\label{sddecomp}
F=F^{+}+F^{-}\,, \qquad *F^{\pm}=\pm\,\im\,F^{\pm}\,,
\ee
where $*$ is the Hodge star of the four-dimensional Minkowski metric. In the 2-spinor formalism, the decomposition of a field strength into its SD and ASD parts is given by
\be\label{fsdasd}
F_{\alpha\dot\alpha\beta\dot\beta} = \eps_{\alpha\beta}\,\tilde F_{\dot\alpha\dot\beta} + \eps_{\dot\alpha\dot\beta}\,F_{\alpha\beta}\,,
\ee
with the SD field strength encoded by $\tilde{F}_{\dot\alpha\dot\beta}=\tilde{F}_{(\dot\alpha\dot\beta)}$ and ASD field strength encoded by $F_{\alpha\beta}=F_{(\alpha\beta)}$. Thus, the self-duality condition $F^-=0$ is $F_{\alpha\beta}=0$ in the 2-spinor formalism.

\medskip

The conformal null infinity of Minkowski space, $\scri$, is obtained in the usual fashion by conformal compactification. To preserve Lorentz invariance, we work with retarded coordinates $(u,r,\lambda _{\alpha},\bar{\lambda }_{\dot\alpha})$, where $\lambda _{\alpha}=(\lambda _0,\,\lambda _1)$ are holomorphic homogeneous coordinates on $\P^1$, subject to the equivalence relation
\be\label{scri1}
(u,r,\,\lambda _{\alpha},\,\bar{\lambda }_{\dot\alpha})\sim (|b|^2\, u,\,|b|^{-2}\,r,\,b\,\lambda _{\alpha},\,\bar{b}\,\bar{\lambda }_{\dot\alpha})\,,\qquad \forall b\in\C^*\,.
\ee
The standard non-homogeneous retarded coordinates $(u,r,\zeta,\bar{\zeta})$ are recovered by choosing a constant future-pointing time-like vector $t^{\alpha\dot\alpha}=\diag\,(1/\sqrt2,1/\sqrt2)$ obeying $t^2=1$, and defining $\hat{\lambda }^{\alpha}:=\sqrt2\,t^{\alpha}{}_{\dot\alpha}\,\bar{\lambda }^{\dot\alpha}$. Rescaling so that $\langle\lambda \,\hat\lambda \rangle=\sqrt2$ gives the original $(u,r)$ and 
\begin{equation*}
\lambda^\alpha=\frac{2^{1/4}}{\sqrt{1+|\zeta|^2}}\,(1,\,\zeta)\,.
\end{equation*}
With such choices, 
\begin{equation*}
x^{\alpha\dot\alpha}=u\,t^{\alpha\dot\alpha}+r\,\lambda^{\alpha}\,\bar{\lambda}^{\dot\alpha}\,,
\end{equation*}
so that $\lambda_\alpha\bar\lambda_{\dot\alpha}$ is the null vector aligned along the outward going light cones from the origin.

Conformal compactification is achieved through the inversion $R=r^{-1}$, with future null infinity $\scri^+$ brought to the finite hypersurface $R=0$ by a conformal rescaling given by
\be\label{res-metric}
\d\hat s^2:=R^2\,\d s^2=R^2\,\d u^2-2\,\d u\,\d R - \frac{4\,\d\zeta\, \d\bar\zeta}{(1+|\zeta|^2)^2}\, .
\ee
The natural conformal equivalence class of degenerate metrics on $\scri^+$ is then written in homogeneous coordinates as
\be\label{scrimet}
\d s^{2}_{\scri^+}=0\times \d u^2+\D\lambda \,\D\bar{\lambda }\,,\qquad \mbox{ where } \qquad\D\lambda :=\langle\lambda \,\d\lambda\rangle:=\lambda ^{\alpha}\,\d\lambda _{\alpha}\,,
\ee
with $(u,\lambda_{\alpha},\bar\lambda_{\dot\alpha})$ the projective Bondi coordinates on $\scri^+\cong\R\times S^2$, $(\lambda,\bar{\lambda})$ being homogeneous coordinates on $S^2$ and $u$ the coordinate on $\R$~\cite{Penrose:1962ij,Penrose:1964ge,Penrose:1965am}.

This projective formulation~\cite{Sparling:1990,Eastwood:1982,Adamo:2014yya,Geyer:2014lca,Adamo:2015fwa} gives rise to the line bundles $\cO(p,q)\rightarrow\scri^+$ whose sections are represented by homogeneous functions $f(u,\lambda ,\bar{\lambda})$ obeying $f(|b|^2 u, b\lambda ,\bar{b}\bar{\lambda })=b^{p}\,\bar{b}^{q}\,f(u,\lambda ,\bar{\lambda })$ for any $b\in\C^*$. The metric \eqref{scrimet} takes values in $\cO(2,2)$, so to fix a conformal scale -- for instance, to give the round metric on $S^2$ -- one must divide by $\la\lambda\,\hat{\lambda}\ra^2$. More generally, functions with non-trivial spin and conformal weight $(s,w)$ on $\scri^+$ in the standard (non-projective) description correspond to sections of $\cO(p,q)$ with $s=\frac{p-q}{2}$ and $w=\frac{p+q}{2}$.  

Translations by $a^{\alpha\dot\alpha}$ on space-time act at $\scri^+$ via $u\rightarrow u+a^{\alpha\dot\alpha}\lambda_\alpha\bar\lambda_{\dot\alpha}$.  Thus $\scri^+$ is the total space of an affine bundle $\cO_{\R}(1,1)\rightarrow\P^1$, whose sections are real-valued functions $f(\lambda,\bar{\lambda})$ which obey $f(b\lambda,\bar{b}\bar{\lambda})=|b|^2 f(\lambda,\bar{\lambda})$. In curved space-time, the choice of zero section is acted on more generally by supertranslations $u\rightarrow u+f$ where $f$ is any section of $\cO_{\R}(1,1)$; thus, there is no preferred choice of zero-section for the $u$-coordinate. 

Points $x\in\M$ are realized at $\scri^+$ by their \emph{light cone cuts}: the intersection of the light cone from $x$ with $\scri^+$. To describe this cut, parametrize the light cone of $x$ by $x^{\alpha\dot\alpha}+ r\,\lambda^\alpha\bar\lambda^{\dot\alpha}$ with $\la\lambda \,\hat\lambda\ra=\sqrt2$ and identify it with $u\,t^{\alpha\dot\alpha}+ r'\,\lambda^\alpha\bar\lambda^{\dot\alpha}$ for some $u,r'$. Contracting with $\lambda_\alpha\bar\lambda_{\dot\alpha}$ gives 
\begin{equation}\label{lcc}
u=x^{\alpha\dot\alpha}\,\lambda_\alpha\,\bar \lambda_{\dot\alpha}\,,
\end{equation}
as the light cone cut of $x$ at $\scri^+$.

Although our focus here and elsewhere will be $\scri^+$, identical considerations apply to $\scri^-$. Indeed, the projective description of $\scri^-$ has fibre coordinate on the total space of $\cO_{\R}(1,1)\rightarrow\P^1$ given by the advanced time coordinate $v=t+r$ rather than $u$. 

\medskip


\paragraph{Radiative gauge fields:} A linear, negative helicity spin-1 field is described by its linearised ASD field strength $\phi_{\alpha\beta}(x)$, which obeys the zero-rest-mass equation $\partial^{\alpha\dot\alpha}\phi_{\alpha\beta}=0$. This field `peels' as $r\rightarrow\infty$ according to~\cite{Penrose:1965am,Penrose:1986uia}
\begin{equation*}
\phi_{\alpha\beta}=\frac{\phi_2}{r}\, \lambda_{\alpha}\, \lambda_{\beta} +O(1/r^2)\,,
\end{equation*}
where $\lambda_\alpha$ is aligned and covariantly constant along the light rays of constant $(u,\zeta,\bar\zeta)$ (i.e., with tangent $\lambda_{\alpha}\bar \lambda_{\dot\alpha}$). The object $\phi_2=\phi_{2}(u,\lambda,\bar{\lambda})$ is the radiation data, defining a section of $\cO(-3,-1)$ on $\scri$ in the homogeneous formalism.
 
Assuming suitable globality, it is straightforward to see that linear spin-1 fields are determined by their radiation data at $\scri$ by means of the Kirchoff-d'Adh\'emar integral formulae adapted to $\scri$ in the unphysical (i.e., conformally rescaled) space-time. This is an integral over the light-cone cut~\cite{Penrose:1980yx,Penrose:1984uia,Penrose:1986uia}
\begin{equation}\label{K-dA}
\phi_{\alpha\beta}(x)= \int_{\P^1} \lambda_{\alpha}\,\lambda_{\beta}\,\D\lambda\wedge\D\bar{\lambda}\,\left. \frac{\p\phi_2}{\p u}\right|_{u=x^{\alpha\dot\alpha}\lambda_{\alpha}\bar\lambda_{\dot\alpha}}\,.
\end{equation}
It is easy to see by differentiating under the integral sign that this solves the zero-rest-mass equation, and corresponds to the given radiation data when the field is defined and    differentiable over $\scri^+$ including the vertex $i^+$ (i.e., future time-like infinity)~\cite{Penrose:1984uia}. The positive helicity spin-1 field $\bar{\phi}_{\dot\alpha\dot\beta}$ is given by complex conjugation.

These assumptions do not apply in the presence of sources (e.g., when the field has Coulombic parts), and these are therefore not encoded in $\phi_2$. This motivates our definition of radiative gauge fields:
\begin{defn}
A \emph{radiative gauge field} in four-dimensional Minkowski space-time is one that is completely characterized by its radiation data, so that $\phi_2$ is the free characteristic data for the field.
\end{defn}
This definition also applies to non-linear gauge fields, where the gauge potential is a more natural starting point than the radiation field itself.

\medskip


\paragraph{Self-dual radiative gauge fields:} Let $A$ be a gauge field in four-dimensional Minkowski space-time, viewed as a 1-form valued in the Lie algebra $\mathfrak{g}$ of a compact gauge group. In retarded Bondi coordinates, an asymptotically flat gauge field in temporal gauge $A_u=0$ can be restricted to $\scri^+$ and pulled back to projective coordinates in the conformally compactified space-time as~\cite{vanderBurg:1969,Newman:1978ze,Strominger:2013lka,Barnich:2013sxa}:
\be\label{AFgt1h}
A|_{\scri^+}=\cA^{0}(u,\lambda,\bar{\lambda})\,\D\lambda+\bar\cA^0(u,\lambda ,\bar{\lambda})\,\D\bar\lambda\,,
\ee
where $\cA^0$ now takes values in $\cO(-2,0)\otimes\mathfrak{g}$ and $\bar{\cA}^0(u,\lambda ,\bar{\lambda })$ in $\cO(0,-2)\otimes\mathfrak{g}$. Thus, $\cA^0$ and $\bar{\cA}^0$ have spin weights $-1$ and $+1$, respectively, and both have conformal weight $-1$.

For the gauge field \eqref{AFgt1h}, the 2-form field strength $F$ at $\scri^+$ has leading (in the conformally rescaled space-time) SD and ASD parts (cf., \cite{Newman:1978ze,Goldberg:1979wt,Newman:1980fr})
\be\label{sddecomp2}
F^{+}|_{\scri^+}=\partial_{u}\bar{\cA}^{0}\,\d u\wedge\D\bar{\lambda}\,, \qquad F^{-}|_{\scri^+}=\partial_{u}\cA^{0}\,\d u\wedge\D\lambda\,,
\ee
so $\cA^0$, $\bar{\cA}^0$ respectively determine the leading asymptotic ASD and SD field strengths, sometimes known as the \emph{broadcasting function} of the gauge field\footnote{In the Newman-Penrose formalism~\cite{Newman:1961qr}, the broadcasting function corresponds to the coefficient of the leading falloff for the gauge field strength component $\Phi_2$.}, to be
\be\label{gtBroad}
\phi_2(u,\lambda ,\bar{\lambda }):=\frac{\partial\cA^{0}}{\partial u}(u,\lambda ,\bar{\lambda })\,,
\ee
with values in $\cO(-3,-1)\otimes\mathfrak{g}$. The broadcasting function, and thus $\cA^0$, actually determines the free radiative data for the fully \emph{non-linear} gauge field~\cite{vanderBurg:1969,Exton:1969im,Ashtekar:1981bq} by \eqref{K-dA} with suitable regularity assumptions. However, we also wish to include plane waves in the class of radiative fields, as they are essentially pure radiation. Thus, we allow for $\cA^0$ to have isolated singular points on the $S^2$ at $\scri^+$ (but generally extended in $u$); such gauge fields will be `almost everywhere' asymptotically flat.

For Lorentz-real gauge fields, there are no purely self-dual (SD) or chiral radiation fields since $\cA^0$ and $\bar{\cA}^0$ are complex conjugates. However, complex gauge fields have independent data $\cA^0$ and $\tilde{\cA}^0$, so it is possible to define a purely SD radiative gauge field: 
\begin{defn}
A \emph{self-dual radiative gauge field} is a complex radiative gauge field that is asymptotically flat almost everywhere and determined by its radiative data $\tilde\cA^0$, with $\cA^0=0$.
\end{defn}

\paragraph{\textit{Example: Self-dual gauge theory plane waves}} An important example of a SD radiative gauge field -- to which we repeatedly return -- is that of a self-dual plane wave (SDPW). This is a highly-symmetric example, being invariant under translations orthogonal to a preferred null direction $n^{\alpha\dot\alpha}=\iota^{\alpha}\,\tilde{\iota}^{\dot\alpha}$, and is furthermore valued in a Cartan subalgebra $\mathfrak{h}\subset\mathfrak{g}$~\cite{Trautman:1980bj,Adamo:2017nia}. 
In lightfront coordinates 
\be\label{coord}
x^{\alpha\dot\alpha} \equiv \begin{pmatrix}x^+&&\tilde z\\z&&x^- \end{pmatrix} \,,
\ee
the metric on $\M$ becomes 
\be\label{eta}
\d s^2 =2\,\left(\d x^+\,\d x^- - \d z\,\d\tilde z\right)\,.
\ee
With $n\cdot x= x^-$ the SDPW gauge field may be written
\be\label{sdpwym}
A = -f(x^-)\,\d\tilde z \,,
\ee
where $f(x^-)$ is an arbitrary function of the lightfront coordinate $x^-$, taking values in the Cartan subalgebra $\mathfrak{h}$. Choosing a spin frame so that $o_\alpha\tilde\iota_{\dot\alpha}\,\d x^{\alpha\dot\alpha}=\d \tilde z$ and $n^{\alpha\dot\alpha}=\iota^\alpha\tilde\iota^{\dot\alpha}$ (i.e., $\la\iota\,o\ra=1=[\tilde{\iota}\,\tilde{o}]$), the field strength is self-dual 
\be\label{sdpwf}
\tilde F_{\dot\alpha\dot\beta} = \dot f\,\tilde\iota_{\dot\alpha}\tilde\iota_{\dot\beta}\,,\qquad F_{\alpha\beta} = 0\,,
\ee
with $\dot f(x^-)\equiv\p_-f(x^-)$. The SDPW gauge field \eqref{sdpwym} is clearly invariant under translations of $x^+$, $z$ and $\tilde z$ but it also has two other (less obvious) null rotation symmetries that together form a five-dimensional Heisenberg algebra whose centre is $n$ (cf., \cite{Heinzl:2017blq,Adamo:2017nia}). When $\dot f$ and hence the field strength is compactly supported in $x^-$, the SDPW is an example of a \emph{sandwich} plane wave, which admits a well-defined S-matrix~\cite{Schwinger:1951nm,Adamo:2017nia}; we assume this sandwich property from now on.

These solutions do not satisfy standard asymptotically flat fall-off conditions as $r\rightarrow \infty$. On one hand, the field is compactly supported along every light ray tangent to $\lambda_\alpha\bar\lambda_{\dot\alpha}$ when $\la\lambda\,\iota\ra\neq 0 $. On the other hand, it is constant and has no fall-off along light rays with $\lambda_\alpha=\iota_\alpha$. We can nevertheless make sense of the radiation data as a distribution supported at $\la\lambda\,\iota\ra=0$. In the projective coordinates for $\scri^{+}$ the free radiative data is given by
\be\label{sdpwA0}
\tilde\cA^0(u,\lambda ,\bar{\lambda})\,\D\bar\lambda=\mathcal{F}\left(\frac{u}{\la\lambda \,o\ra\,[\bar{\lambda }\,\tilde{o}]}\right)\,\bar \delta \left(\frac{\la \iota\,\lambda \ra}{\la o\,\lambda\ra}\right)\,,
\ee
where the Cartan-valued function\footnote{By `Cartan-valued function' we always mean a function which takes values in the Cartan subalgebra $\mathfrak{h}\subset\mathfrak{g}$.} $\mathcal{F}$ is the first antiderivative of $f$ ($\dot{\mathcal{F}}=f$), and the holomorphic delta function (technically, a $(0,1)$-distribution) is defined as 
\begin{equation*}
\bar\delta (z):=\frac{1}{2\pi\im}\,\bar\p \frac{1}{z} =\delta(\Re z)\,\delta(\Im z) \, \d\bar z\,,
\end{equation*}
for any complex variable $z$. Thus, $\tilde{\cA}^0$ has a double delta function singularity at the generator of $\scri^+$ where $\la \lambda\, \iota\ra=0$ and so is only almost everywhere asymptotically flat. The SDPW broadcasting function is therefore\footnote{That the curvature component $\tilde \phi_2$ at $\scri^+$ should be identified with the gauge dependent $f$ is perhaps a puzzle. This is resolved by the fact that the radiative assumption gives the preferred choice for $f$ that arises by integrating the gauge invariant $\dot f $ from $i^+$ so that $i^+$ is regular.}  
\be\label{sdpwBroad}
\tilde\phi_2(u,\lambda ,\bar{\lambda })\,\D\bar\lambda= \frac{f\left(\frac{u}{\la\lambda \,o\ra\,[\bar{\lambda }\,\tilde{o}]}\right)}{\la\lambda \,o\ra\,[\bar{\lambda }\,\tilde{o}]}\,\bar \delta \left(\frac{\la \iota\,\lambda\ra}{\la o\,\lambda\ra}\right)\,.
\ee
These identifications can be verified by means of \eqref{K-dA}.


\subsection{Twistor theory of SD radiative gauge fields}
\label{YMTwTh}

The twistor space $\PT$ of complexified Minkowski space $\M\cong\C^4$ is an open subset of $\P^3$. Using homogeneous coordinates $Z^{A}=(\mu^{\dot\alpha},\lambda_{\alpha})$, define\footnote{We follow the conventions of~\cite{Adamo:2017qyl}.}
\be\label{PTmink}
\PT=\P^3\setminus\left\{Z^A\,|\,\lambda_{\alpha}=0\right\}\,.
\ee 
Points $x\in\M$ correspond to holomorphic, linearly embedded Riemann spheres $X\cong\P^1$ in $\PT$~\cite{Penrose:1967wn} given by the incidence relation
\begin{equation}\label{inc}
\mu^{\dot\alpha}=x^{\alpha\dot\alpha}\lambda_\alpha\, .
\end{equation}
For such finite $x^{\alpha\dot\alpha}$, the twistor variables $\lambda_{\alpha}$ are identified with the homogeneous coordinate $\lambda _{\alpha}$ on each twistor `line' $X\cong\P^1$. It is easy to see that two twistor lines $X,Y\subset\PT$ intersect if and only if the corresponding points $x,y\in\M$ are null separated. Conversely, holding $Z^A$ constant, \eqref{inc} corresponds to a totally null ASD two-plane in complexfied Minkowski space $\M$.

Selecting a real slice of $\M$ with Euclidean $\R^4$ or split $\R^{2,2}$ signature gives rise to a complex conjugation on $\PT$ that is quaternionic for Euclidean signature~\cite{Atiyah:1978wi,Woodhouse:1985id}, or that fixes the $\RP^3$ on which all components of $Z^A$ are real for split signature. Our focus in this paper will be on Lorentz signature $\R^{1,3}$ where the reality structure induces a pseudo-Hermitian structure of signature $(2,2)$ on $\PT$ given by 
 \begin{equation}
 Z\cdot \bar Z:=\mu^{\dot\alpha}\bar \lambda_{\dot\alpha}-\lambda_\alpha\bar\mu^\alpha\, .
 \end{equation}
Points of real Minkowski space correspond to lines $X$ that lie in $\PN=\{Z\in \PT|Z\cdot \bar Z=0\}$, which conversely is the space of real light rays in $\R^{1,3}$.

Since each light ray in $\R^{1,3}$ makes it out to $\scri^+$, there is a natural projection given by 
\begin{equation}
 p:\PN\rightarrow \scri^+\, , \qquad (\mu^{\dot\alpha},\lambda_{\alpha})\mapsto (u=\mu^{\dot\alpha}\bar\lambda_{\dot\alpha},\,\lambda_\alpha,\,\bar\lambda_{\dot\alpha})\, , \label{scri-inc}
\end{equation}
where the reality of $u$ follows from $Z\cdot\bar Z=0$. It follows from the formula in \eqref{scri-inc} that $p$ naturally extends to
\be\label{scri-inc-C}
p:\PT\rightarrow \scri^+_\C\, ,
\ee
where $\scri^+_\C$ is a partial complexification of $\scri^+$ obtained by taking $u\in \C$ but retaining $\tilde\lambda_{\dot\alpha}=\bar\lambda_{\dot\alpha}$ inside the full complexification. Thus $p$ is holomorphic in $\mu^{\dot\alpha}$ but of course not in $\lambda_\alpha$. The map $p$ projects $X$ given by \eqref{inc} to the light cone cut of $x$ in $\scri^+$ \eqref{lcc}.

Characteristic data on $\scri^+_\C$ can be pulled back using $p$ to $\PT$ (or $\PN$ if it has no analytic extension in $u$). On $\PT$ the Kirchoff-d'Adh\'emar integral formula \eqref{K-dA} can be re-interpreted~\cite{Mason:1986} as a gauge fixed version of the Penrose transform~\cite{Penrose:1969ae,Eastwood:1981jy}. The Penrose transform is the correspondence
\begin{equation*}
H^1(\PT,\cO(-4)) =\left\{ \phi_{\alpha\beta}(x) \mbox{ on } \M\,|\,\partial^{\alpha\dot\alpha} \phi_{\alpha\beta}=0\right\}\, .
\end{equation*}
In the Dolbeault representation of the cohomology group, an element $f\in H^1(\PT,\cO(-4))$ is a $\dbar$-closed $(0,1)$-form on twistor space modulo exact forms of homogeneity weight $-4$. The correspondence operates via the integral formula
\begin{equation}\label{ASD-PT}
\phi_{\alpha\beta}(x)=\int_X \lambda_{\alpha}\,\lambda_{\beta}\, \D\lambda\wedge f|_{X}\,,
\end{equation}
where $f|_{X}$ denotes restriction to the twistor line $X\cong\P^1$ via the incidence relations \eqref{inc}. The Kirchoff-d'Adh\'emar integral formula \eqref{K-dA} follows as a special gauge fixing within the Dolbeault equivalence class via the identification 
\begin{equation}
f=\frac{\partial\phi_2}{\partial u}\, \D\bar \lambda\,,
\end{equation}
which is easily seen to be homogeneous of weight $-4$ in $Z^{A}$ and $\dbar$-closed.

For positive helicity, the Penrose transform reads
\begin{equation*}
H^1(\PT,\cO) =\bigl\{ \tilde\phi_{\dot\alpha\dot\beta}(x) \mbox{ on } \M\,|\,\partial^{\alpha\dot\alpha}\tilde \phi_{\dot\alpha\dot\beta}=0\bigr\}\,,
\end{equation*}
with integral formula
\begin{equation}
\tilde\phi_{\dot\alpha\dot\beta}(x)=\int_X \left.\frac{\p^2f}{\p \mu^{\dot\alpha}\p\mu^{\dot\beta}}\right|_{X}\wedge\D\lambda\,.\label{SD-PT}
\end{equation}
By setting $f= \tilde\cA^0\,\D\bar\lambda$, this can be identified with the conjugated version of \eqref{K-dA}.

\medskip


\paragraph{Twistor description of SD radiative gauge fields:} The positive helicity Penrose transform has nonlinear extensions for SD gauge fields~\cite{Ward:1977ta}, with wide application to the construction of monopoles and instantons, and indeed to the general theory of integrable systems (cf., \cite{Ward:1990vs,Mason:1991rf,Dunajski:2010zz}). For the SD Yang-Mills equations, the relevant construction is:
\begin{thm}[Ward~\cite{Ward:1977ta}]\label{Ward}
There is a one-to-one correspondence between:
\begin{itemize}
 \item SD gauge fields on $\M$, and 
 
 \item holomorphic vector bundles $E\to\PT$ with $E|_X$ topologically trivial for every twistor line $X\cong\P^1$ corresponding to $x\in\M$.
\end{itemize}
\end{thm}
Further restrictions on $E\rightarrow\PT$ give details of the gauge group on space-time and reality structure; for example, for gauge group SU$(N)$ in split signature, $E$ has rank $N$, is equipped with a positive definite real form on the real slice of $\PT$, and $\det E$ is trivial~\cite{Mason:2005qu}. The bundle $E\rightarrow\PT$ is often referred to as the \emph{Ward bundle}.

Theorem~\ref{Ward} can also be understood in terms of radiative data at $\scri^+$ via a Dolbeault presentation. This expresses the complex structure on the Ward bundle $E\rightarrow\PT$ in terms of a $\bar\p$-operator $\Dbar$ taking sections of $E$ to $(0,1)$-forms that satisfies $\Dbar^2=0$. SD Yang-Mills is conformally invariant and the bundle and connection naturally extend to $\scri^+$ as a bundle with the connection given in \eqref{AFgt1h}. The Ward bundle for a SD radiative gauge field then arises simply by pulling back this bundle with $\dbar$-operator arising from \eqref{AFgt1h} to $\PT$~\cite{Sparling:1990,Newman:1978ze,Newman:1980fr}:
\begin{equation}
\Dbar=\dbar + \sa\,, \qquad \sa:=\tilde\cA^0(\mu^{\dot\alpha}\bar\lambda_{\dot\alpha},\lambda,\bar\lambda)\,\D\bar \lambda\, ,\label{radiativeym}
\end{equation}
with the $(0,1)$-form $\sa$ valued in $\End\,E$. That $\Dbar^2=0$ follows trivially, as the only anti-holomorphic dependence of $\sa$ is in $\bar\lambda$, and $\D\bar\lambda\wedge \D\bar\lambda=0$.

\paragraph{Proof of Theorem \ref{Ward}:} In the non-abelian case, we can work on $\scri^+$ following~\cite{Sparling:1990,Newman:1978ze,Newman:1980fr}. 
Upon restriction to any line $X\subset\PT$, it follows from the assumptions\footnote{In fact with arbitrary data, this will generically be the case~\cite{Jiang:2008xw,Mason:2010yk,Bullimore:2011ni}, although see \cite{Sparling:1990} for exceptions.} in Theorem \ref{Ward} that $E$ is holomorphically trivial, which is the statement that we can find a frame $\sH(x,\lambda,\bar\lambda):E|_X\rightarrow \C^{r}$ of $E$, for $E$ of rank $r$, so that $\Dbar|_X \sH=0$. This relation is sometimes referred to as the `Sparling equation,' and by Liouville's theorem any solution will be unique up to $\sH\rightarrow \sH g(x)$ for some matrix function $g$ independent of $\lambda$. Given such an $\sH$, it follows that  
\begin{equation*}
\dbar|_{X}(\sH^{-1}\lambda^\alpha\p_{\alpha\dot\alpha} \sH)=0\,,
\end{equation*}
since $\lambda^\alpha\p_{\alpha\dot\alpha}(\sa|_X)=0$. Thus $\sH^{-1}\lambda^\alpha\p_{\alpha\dot\alpha} \sH$ is holomorphic in $\lambda_\alpha$ and of homogeneity degree one, so
\begin{equation}\label{Lax0}
\sH^{-1}\lambda^\alpha\p_{\alpha\dot\alpha} \sH=-\im\,\lambda^\alpha\, A_{\alpha\dot\alpha}(x)\, ,
\end{equation}
for some $A_{\alpha\dot\alpha}$ on $\M$ taking values in the adjoint of the gauge group. Under $\sH\rightarrow \sH g(x)$ it follows that $D_{\alpha\dot\alpha}=\p_{\alpha\dot\alpha}-\im\,A_{\alpha\dot\alpha}$ transforms as a gauge field with $g$ being the gauge transformation. Equation \eqref{Lax0} is equivalent to the Lax pair equations
\begin{equation}\label{Lax}
\lambda^\alpha\, D_{\alpha\dot\alpha}\sH^{-1}=0\, ,
\end{equation}
the integrability of which implies
\begin{equation}
0 = [\lambda^\alpha D_{\alpha\dot\alpha},\,\lambda^\beta D_{\beta\dot\beta}]= \lambda^\alpha\lambda^\beta\, F_{\alpha\beta}\,\varepsilon_{\dot\alpha\dot\beta}\, ,
\end{equation}
for $F_{\alpha\beta}$ the ASD field strength of the gauge connection on $\M$. Thus, integrability of the Lax pair is equivalent to the SD Yang-Mills equations $F_{\alpha\beta}=0$ for the connection $D_{\alpha\dot\alpha}$. \qed

\medskip

In general, it is hard to solve for $\sH$ explicitly. However, when $\tilde\cA^0$ takes values in a Cartan subalgebra $\mathfrak{h}\subset\mathfrak{g}$, the $\Dbar$ in \eqref{radiativeym} defines a cohomology class $\sa\in H_{\dbar}^{0,1}(\PT,\cO\otimes\mathfrak{h})$, and the SD radiative gauge field on $\M$ can be recovered using \eqref{SD-PT}. In this case, $\sa|_{X}\in H_{\dbar}^{0,1}(\P^1,\cO\otimes\mathfrak{h})$; but this cohomology group vanishes, so
\be\label{splitym}
\sa|_{X}=\dbar|_{X} g(x,\lambda)\,,
\ee
for some Cartan-valued $g(x,\lambda)$ of homogeneity zero in $\lambda$. This can be obtained explicitly from a Green's function on the sphere as 
\begin{equation}\label{Greens-fn}
g(x,\lambda)=\frac{1}{2\pi\im}\int_{X}\frac{\D\lambda'}{\la\lambda\,\lambda'\ra}\,\frac{\la o\,\lambda\ra}{\la o\,\lambda'\ra}\;\sa(x,\lambda')=\frac{1}{2\pi\im}\int_{X}\frac{\D\lambda'\wedge\D\bar{\lambda}'}{\la\lambda\,\lambda'\ra}\,\frac{\la o\,\lambda\ra}{\la o\,\lambda'\ra}\;\tilde{\cA}^{0}(x,\lambda')\, ,
\end{equation}
where the choice of the spinor $o_\alpha$ amounts to an arbitrary gauge choice. 

With this we see that $\Dbar\sH|_{X}=0$ is solved by setting
\be\label{Hdef1}
\sH(x,\lambda )=\exp\left[-g(x,\lambda )\right]\,,
\ee
where $\exp(\cdots)$ denotes the exponential map of the gauge group. Now \eqref{Lax0} gives
\begin{equation}\label{descentym}
\begin{split}
\lambda^\alpha\, A_{\alpha\dot\alpha}=-\im\,\lambda^\alpha\p_{\alpha\dot\alpha}g&=-\frac{\la o\,\lambda\ra}{2\,\pi}\int_{X}\frac{\D\lambda'}{\la o\,\lambda'\ra}\;\frac{\partial\sa}{\partial\mu^{\dot\alpha}}(x,\lambda') \\
&=-\frac{\la o\,\lambda\ra}{2\,\pi}\int_{X}\frac{\D\lambda'}{\la o\,\lambda'\ra}\;\bar{\lambda}'_{\dot\alpha}\,\tilde{\phi}_{2}(x,\lambda')\,,
\end{split}
\end{equation}
where $\tilde{\phi}_{2}=\partial_{u}\tilde{\cA}^{0}$ is the SD broadcasting function. From this one obtains the integral formula for $A$ itself:
\begin{equation}\label{int-pot}
\begin{split}
A_{\alpha\dot\alpha}(x)&=\frac{o_{\alpha}}{2\pi} \int_X \frac{\D\lambda}{\la o\,\lambda\ra}\,\left.\frac{\partial\sa}{\partial\mu^{\dot\alpha}}\right|_{X} \\
 &=\frac{o_{\alpha}}{2\pi}\int_{X}\frac{\D\lambda\wedge\D\bar{\lambda}}{\la o\,\lambda\ra}\,\bar{\lambda}_{\dot\alpha}\,\tilde{\phi}_{2}|_{X}\,, 
\end{split}
\end{equation}
expressed equivalently in terms of the complex structure on the Ward bundle or the radiative data.

\medskip


\paragraph{Background coupled gluons:}  A linearised gluon field $a_{\alpha\dot\alpha}(x)$ propagating in a background gauge field $A_{\alpha\dot\alpha}$ has a linearised field strength $f_{ab}=D_{[a}a_{b]}$ which can be decomposed into its SD and ASD parts: $\tilde{f}_{\dot\alpha\dot\beta}$ and $f_{\alpha\beta}$, respectively. However, on a non-trivial background there is no longer a natural global decomposition $a_b=a^+_b+a_b^-$ of the potential into SD and ASD parts, as an infinitesimal gauge transformation $a_a=D_a\gamma$ leads to $f_{\alpha\beta}=[F_{\alpha\beta},\gamma]$ where $F$ is the curvature of the background field, so such a decomposition cannot be gauge invariant. When the background gauge field is SD, it \emph{is} gauge invariant to characterise a positive helicity gluon by $f_{\alpha\beta}=0$. Although it is not possible to write down a gauge invariant potential for a negative helicity gluon, it can nevertheless be characterised asymmetrically by a curvature 2-form $f_{\alpha\beta}$ satisfying $D^{\alpha\dot\alpha}f_{\alpha\beta}=0$. 

For a positive helicity gluon (with suitable analyticity) on a SD radiative gauge field background, the Penrose transform represents the gluon in terms of cohomology in twistor space~\cite{Penrose:1969ae,Eastwood:1981jy,Ward:1990vs}:
\be\label{+glpt1}
H^{0,1}_{\Dbar}(\PT,\,\cO\otimes\mathrm{End}\,E)\cong\bigl\{a_{\alpha\dot\alpha} \mbox{ on } \M\,|\,f_{\alpha\beta}:=D^{\dot\alpha}_{(\alpha}a_{\beta)\dot\alpha}=0\bigr\}\,,
\ee
where $H^{0,1}_{\Dbar}$ denotes the Dolbeault cohomology with respect to $\Dbar$. To see how this isomorphism works in the direction from twistor space to space-time, let $a\in H^{0,1}_{\Dbar}(\PT,\cO\otimes\End\,E)$. Upon restriction to the twistor line $X$, $E|_X$ is holomorphically trivial with trivialization $\sH$, so $a|_{X}$ takes values in a trivial bundle on $X$. In this holomorphic frame $a|_X$ becomes $\sH^{-1}\,a|_{X}\,\sH\in H_{\dbar}^{0,1}(\P^1,\,\cO\otimes\mathfrak{g})$, and the triviality of this cohomology group implies that the Green's function used in \eqref{Greens-fn} can be used to find some $j(x,\lambda)$ such that
\be\label{sdsplitym}
\sH^{-1}\,a|_X\,\sH = \dbar|_X j(x,\lambda)\,,
\ee
for $j(x,\lambda )$ valued in the adjoint and of homogeneity zero in $\lambda $.

Acting on both sides of \eqref{sdsplitym} with $\lambda^\alpha D_{\alpha\dot\alpha}$ and using \eqref{Lax} gives $\dbar|_X(\lambda ^\alpha D_{\alpha\dot\alpha}j) = 0$. By Liouville's theorem, this implies the existence of an adjoint-valued gluon field $a_{\alpha\dot\alpha}(x)$ such that
\be\label{sddescentym}
\lambda^\alpha\, D_{\alpha\dot\alpha}j(x,\lambda ) = \im\,\lambda^\alpha\, a_{\alpha\dot\alpha}(x)\,.
\ee
Contracting both sides of \eqref{sddescentym} with $\lambda ^\beta D_\beta^{\dot\alpha}$ (and recalling that the background is SD and $j$ is essentially a perturbation to $\sH$) shows that this is a positive helicity gluon (i.e., $f_{\alpha\beta}=0$), as desired. 

For a negative helicity gluon in a SD radiative background gauge field, the Penrose transform is
\be\label{-glpt1}
H^{0,1}_{\Dbar}(\PT,\,\cO(-4)\otimes\mathrm{End}\,E)\cong\left\{a_{\alpha\dot\alpha} \mbox{ on } \M\,|\,D^{\alpha\dot\alpha}f_{\alpha\beta}=0\right\}\,.
\ee
In this case, for some $b\in H^{0,1}_{\Dbar}(\PT,\cO(-4)\otimes\End\,E)$ the space-time field is given by the integral
\be\label{GPenrose}
f_{\alpha\beta}(x) = \int_X\D\lambda \wedge\lambda _\alpha\,\lambda _\beta\,\sH^{-1}\,b|_X\,\sH\,.
\ee
This integral produces a well-defined adjoint-valued object on $\M$ as $\sH^{-1}b|_X\sH$ takes values in $H^{0,1}(\P^1,\cO(-4)\otimes\mathfrak{g})$. Using \eqref{Lax}, it follows that
\be\label{Bsolves}
D^{\alpha\dot\alpha}f_{\alpha\beta} = \int_X\D\lambda \wedge\lambda _\beta\,\sH^{-1}\left(\lambda _\alpha\p^{\alpha\dot\alpha}\,b|_X\right)\sH = 0\,,
\ee
as required. 

\medskip


\paragraph{Momentum eigenstates on SD backgrounds:} For the computation of scattering amplitudes, it is useful to have cohomological representatives corresponding to momentum eigenstates. That is, what are the twistor representatives corresponding to gluons which have null momentum $k_{\alpha\dot\alpha}=\kappa_{\alpha}\,\tilde{\kappa}_{\dot\alpha}$ at $\scri^{-}_{\C}$ (i.e., before passing through the non-trivial SD radiative background)? Remarkably, the representation \eqref{radiativeym} for the partial connection $\Dbar$ ensures that standard twistor representatives for momentum eigenstates (in a trivial background) provide the answer.

For a positive helicity gluon, consider the representative
\be\label{+glme}
a(Z)=\sT^{\sa}\,\int_{\C^*}\frac{\d s}{s}\,\bar{\delta}^{2}(\kappa-s\,\lambda)\,\e^{\im\,s\,[\mu\,\tilde{\kappa}]}=\sT^{\sa}\,\int_{\C^*}\frac{\d s}{s}\,\bar{\delta}^{2}(\kappa-s\,\lambda)\,\e^{\im\,s\bar s\, u}\,,
\ee
where $\sT^\sa$ is a generator of the adjoint representation of the gauge group and the holomorphic delta function
\begin{equation*}
\bar{\delta}^{2}(\kappa-s\,\lambda):=\frac{1}{(2\pi\im)^2}\,\bigwedge_{\alpha=0,1}\dbar\left(\frac{1}{\kappa_{\alpha}-s\,\lambda_\alpha}\right)\,,
\end{equation*}
has support only where both components of its argument vanish, i.e., where  $\lambda_{\alpha}=s\kappa_\alpha$. The second equality in \eqref{+glme} follows on the support of the delta function only when $\tilde \kappa =\bar\kappa$, but then serves to give $a(Z)$ in terms of its distributional characteristic data for a momentum eigenstate.  It is straightforward to show that $a$ is $\Dbar$-closed as a consequence of the fact that $a$ is proportional to $\D\bar{\lambda}$ and only depends on anti-holomorphic variables through $\bar\lambda$.

To obtain the corresponding gluon wavefunction, one follows \eqref{sdsplitym} -- \eqref{sddescentym}; on a general non-abelian SD radiative background, it is difficult to solve for $\sH$ explicitly, so let us make the simplifying assumption that the background is Cartan-valued. In order to keep track of the colour structure, assume that $\msf{T^a}$ is an eigenstate of the Cartan $\mathfrak{h}$ so that the colour structure of the representative is not deformed when conjugating by the holomorphic frame. Thus $\msf{T^a}$ defines a `charge' $e^\msf{i}$ with respect to the background, defined as the root $e=(e^\msf{i})_{1\leq\msf{i}\leq\text{rank}(\g)}$, with $e^\msf{i}$ defined by $[\msf{t^i},\msf{T^a}]=e^\msf{i}\,\msf{T^a}$ for $\{\msf{t^i}\}$ a basis of the Cartan subalgebra. For any Cartan-valued function $f$, we denote $[f(x^-),\msf{T^a}] = e^\msf{i}\,f^\msf{i}\,\msf{T^a}$ unambiguously as $e\,f\,\msf{T^a}$.

Upon performing the scale integral in \eqref{+glme} and applying \eqref{Hdef1}, this means that
\be\label{sdme1}
\sH^{-1}(x,\lambda )\,a|_{X}\,\sH(x,\lambda )=\sT^{\sa}\,\frac{\la \xi\,\lambda\ra}{\la\xi\,\kappa\ra}\,\bar{\delta}(\la\lambda \,\kappa\ra)\,\e^{\im\,k\cdot x+e\,g(x,\kappa)}\,,
\ee
where $\xi^{\alpha}$ is an arbitrary constant spinor (on which the expression does not depend). From \eqref{sdme1} and \eqref{sdsplitym}, it follows that
\be\label{sdme2}
j(x,\lambda )=\sT^{\sa}\,\frac{\la \xi\,\lambda\ra}{\la\xi\,\kappa\ra\,\la\lambda \,\kappa\ra}\,\e^{\im\,\phi(x)}\,,
\ee
where the function 
\be\label{gHamJac}
\phi(x):=k\cdot x-\im\,e\,g(x,\kappa)\,,
\ee
is easily seen to solve the Hamilton-Jacobi equation $(\p^{a}\phi-e\,A^a)\,(\p_{a}\phi-e\,A_a)=0$.

To determine the positive helicity gluon field itself, it helps to rewrite the relation \eqref{descentym} as
\be\label{sdsplit2}
 \partial_{\alpha\dot\alpha}g(x,\lambda )=\im\,A_{\alpha\dot\alpha}(x)+\im\,\lambda _{\alpha}\,g_{\dot\alpha}(x,\lambda )\,,
\ee
for $g_{\dot\alpha}(x,\lambda )$ two Cartan valued functions which are homogeneous of degree $-1$ in $\lambda $. Using the Schouten identity, these can be explicitly found from \eqref{Greens-fn} and \eqref{int-pot} to be
\be\label{gdalexp}
g_{\dot\alpha}(x,\lambda) = -\frac{1}{2\pi}\int_X\frac{\D\lambda'}{\la\lambda\,\lambda'\ra}\;\frac{\p\sa}{\p\mu^{\dal}}(x,\lambda') = -\frac{1}{2\pi}\int_X\frac{\D\lambda'\wedge\D\bar\lambda'}{\la\lambda\,\lambda'\ra}\;\bar\lambda'_{\dal}\,\tilde\phi_2(x,\lambda')\,.
\ee
From \eqref{sdme2} and \eqref{sdsplit2} it then follows that
\be\label{Aplus}
a^{(+)}_{\alpha\dot\alpha} = \msf{T^a}\;\frac{\xi_\alpha\,\tilde K_{\dot\alpha}(x)}{\la\xi\,\kappa\ra}\,\e^{\im\,\phi(x)}\,,
\ee
where the spinor
\be\label{dspinor}
\tilde{K}_{\dot\alpha}(x):=\tilde{\kappa}_{\dot\alpha}+e\,g_{\dot\alpha}(x,\kappa)\,,
\ee
defines the on-shell null momentum of the gluon field $K_{\alpha\dot\alpha}(x) := \p_{\al\dal}\phi - e\,A_{\al\dal}=\kappa_{\alpha}\,\tilde{K}_{\dot\alpha}(x)$ at any point; self-duality of the background means that only the dotted momentum spinor is `dressed' by the background.

For a negative helicity gluon with momentum $\kappa_{\alpha}\tilde{\kappa}_{\dot\alpha}$ at $\scri^{-}_{\C}$, take the representative
\be\label{-glme}
b(Z)=\sT^{\sa}\,\int_{\C^*}\d s\,s^{3}\,\bar{\delta}^{2}(\kappa-s\,\lambda)\,\e^{\im\,s\,[\mu\,\tilde{k}]}\,.
\ee
It is easy to show that $b\in H^{0,1}_{\Dbar}(\PT,\cO(-4)\otimes\End\,E)$, and feeding this into the integral formula \eqref{GPenrose} gives 
\be\label{GPenrose1}
\begin{split}
f^{(-)}_{\alpha\beta}(x) &= \int_X\D\lambda \wedge\lambda _\alpha\,\lambda _\beta\,\sH^{-1}\,b|_X\,\sH \\
 & = \mathsf{T^a}\,\kappa_{\alpha}\,\kappa_{\beta}\,\e^{\im\,\phi(x)}\,.
\end{split}
\ee
As expected by the self-duality of the background, this differs from the linearised field strength of a negative helicity gluon in a trivial background only by the appearance of $\phi(x)$ in the exponential.

\medskip


\paragraph{\textit{Example: SDPWs}} For the special case of the self-dual plane wave (SDPW) given by \eqref{sdpwym}, the partial connection on the associated Ward bundle $E\rightarrow\PT$ is given by
\be\label{twistorsdpwym}
\sa = 2\,\pi\,\int_{\C^*}\frac{\d s}{s}\,\bar\delta^2(\iota-s\,\lambda)\,\F(s[\mu\,\tilde\iota])\,,\qquad\text{where }\F(x^-) := \int^{x^-}f(t)\,\d t\,,
\ee
from which it follows that
\be\label{splitsdpwym}
g(x,\lambda ) = -\im\,\frac{\la o\,\lambda \ra}{\la\iota\,\lambda\ra}\,\F(x^-)\,.
\ee
Plugging this into \eqref{int-pot}, we immediately recover the SDPW gauge potential $A_{\alpha\dot\alpha}(x)=-f(x^-)\, o_{\alpha}\tilde{\iota}_{\dot\alpha}$, and from \eqref{sdsplit2},
\be\label{gsdpw}
g_{\dot\alpha}(x,\lambda )=\frac{\tilde{\iota}_{\dot\alpha}}{\la\iota\,\lambda\ra}\,f(x^-)\,,
\ee
upon using the Schouten identity or equivalently \eqref{gdalexp}. 

In lightfront coordinates \eqref{coord}, the components of an on-shell 4-momentum are
\be\label{lc4mom}
k_{\alpha\dot\alpha}=\kappa_{\alpha}\,\tilde{\kappa}_{\dot\alpha}=\left(\begin{array}{cc}
																									k_+ & \tilde{k} \\
																									k & \frac{k\,\tilde k}{k_+}
																									\end{array}\right)
																									=\begin{pmatrix}\sqrt{k_+}\\\frac{k}{\sqrt{k_+}}\end{pmatrix}_\al\begin{pmatrix}\sqrt{k_+}&\frac{\tilde k}{\sqrt{k_+}}\end{pmatrix}_{\dal}\,;
\ee
and combined with \eqref{splitsdpwym} this yields the associated solution to the Hamilton-Jacobi equation \eqref{gHamJac}:
\be\label{sdpwHJ}
\phi(x)=k\cdot x +e\,\frac{k}{k_{+}}\,\cF(x^-)\,.
\ee
As expected, this is the chiral truncation of the solution to the Hamilton-Jacobi equation in a general plane wave background~\cite{Wolkow:1935zz,Seipt:2017ckc,Adamo:2017nia}. The dressed on-shell momentum at any lightfront time
\be\label{sdpwdressed}
K_{\alpha\dot\alpha}(x^-) = \kappa_{\alpha}\,\tilde{K}_{\dot\alpha}(x^-)\,, \qquad \tilde{K}_{\dot\alpha}=\tilde{\kappa}_{\dot\alpha}+\frac{e\,f(x^-)}{\la\iota\,\kappa\ra}\,\tilde{\iota}_{\dot\alpha}\,,
\ee
also matches the chiral projection of a dressed gluon momentum in a generic plane wave background to the SD sector~\cite{Adamo:2019zmk,Adamo:2020qru}.

Making use of the null symmetry of the SDPW background, the lightfront gauge condition $n\cdot a=\iota^{\alpha}\tilde{\iota}^{\dot\alpha}\,a_{\alpha\dot\alpha}=0$ can be imposed on both positive and negative helicity gluon fields. In this positive helicity case of \eqref{Aplus}, this simply fixes $\xi_{\alpha}=\iota_{\alpha}$, while in the negative helicity case it allows explicit determination of the gluon wavefunction:
\be\label{Anegative}
a^{(-)}_{\alpha\dot\alpha}=\mathsf{T^a}\,\frac{\kappa_{\alpha}\,\tilde{\iota}_{\dot\alpha}}{[\tilde{\iota}\,\tilde{\kappa}]}\,\e^{\im\,\phi(x)}\,.
\ee
This transverse-lightfront gauge is the only gauge for which the negative helicity gluon has no SD linearised field strength.


\section{MHV scattering on gauge theory backgrounds}
\label{SECT:MHVgauge}

We now turn to tree-level MHV scattering of gluons in a SD radiative gauge field background. The space-time generating functional for these amplitudes is lifted to twistor space and perturbatively expanded to yield an all-multiplicity expression for MHV scattering which displays much of the simplicity of its flat background cousin, the Parke-Taylor formula~\cite{Parke:1986gb}.


\subsection{The MHV generating functional}

The chiral formulation of gauge theory in (complexified) Minkowski space $\M$ is given by the Chalmers-Siegel action~\cite{Chalmers:1996rq}
\be\label{CSaction}
S[A,B] = \int_{\M}\d^4 x\;\tr\left(F_{\alpha\beta}\,B^{\alpha\beta} + \frac{\rg^2}{4}\,B_{\alpha\beta}\,B^{\alpha\beta}\right)\,,
\ee
where $F_{\alpha\beta}$ is the ASD part of the field strength of the gauge field $A$, $B_{\alpha\beta}$ is an auxiliary ASD 2-form valued in the adjoint of the gauge group, and $\rg$ is the dimensionless coupling constant. The field equations are
\be\label{CSeom}
F_{\alpha\beta} = -\frac{\rg^2}{2}\,B_{\alpha\beta}\,,\qquad D_{\alpha\dot\alpha}B^{\alpha\beta} = 0\,,
\ee
so $B_{\alpha\beta}$ can be integrated out, resulting in a Lagrangian which differs from the standard Yang-Mills Lagrangian by a multiple of the topological term $\tr(F\wedge F)$. Thus, the Chalmers-Siegel action is perturbatively equivalent to Yang-Mills theory in Minkowski space, and will produce the same gluon amplitudes. Furthermore, it is apparent from \eqref{CSeom} that in this formulation, $\rg=0$ corresponds to the SD sector (i.e., $F_{\alpha\beta}=0$). Thus, small $\rg\ll1$ perturbation theory corresponds to a perturbative expansion around the integrable SD sector, with the auxiliary field $B_{\alpha\beta}$ encoding ASD fluctuations on a SD background.

This formalism is particularly useful for giving a geometric perspective on MHV scattering amplitudes. The MHV helicity configuration (on any perturbative background) involves two negative helicity gluons and an arbitrary number of positive helicity gluons. When the perturbative background is trivial or a SD radiative gauge field, this configuration can be realized as a two-point function of the negative helicity gluons on a SD background~\cite{Mason:2008jy}. Expanding this composite background into a non-linear SD radiative gauge field and a superposition of positive helicity external gluons then recovers the desired MHV amplitude.

The chiral Chalmers-Siegel formulation is well-adapted to this setup. Let $\mathcal{R}$ denote the infinite-dimensional phase space of fields $(A,B)\in \Omega^1(\mathfrak{g})\oplus \Omega^{2}_{-}(\mathfrak{g})$ which solve the field equations \eqref{CSeom}, and $\mathcal{P}$ the Yang-Mills phase space obtained as the quotient of $\mathcal{R}$ by gauge transformations. Let $(A,0)\in\mathcal{R}$ be a SD solution that can be perturbed to define a superposition of a SD radiative gauge field and an arbitrary number of positive helicity gluons on that SD background. The boundary term of \eqref{CSaction} defines a symplectic form on $\mathcal{P}$:
\be\label{gaugesym}
\omega=\frac{1}{\rg^2}\int_{C}\tr\left(\delta A\wedge\delta B\right)\,,
\ee
where $C\subset\M$ is a three-dimensional Cauchy surface and $\delta$ is the exterior derivative on $\mathcal{R}$. It is standard that $\omega$ is independent of the choice of $C$ and descends to become nondegenerate on $\mathcal{P}$ after quotienting by gauge transformations.\footnote{With a definition of positive frequency for the perturbations $\mathscr{A}=(\delta A,\delta B)$, say at $\scri^+$, this defines a positive definite norm by $\la \mathscr{A}_1|\mathscr{A}_2\ra=\im\,\omega(\bar{\mathscr{A}}_1,\mathscr{A}_2)$. }

Let  $U=T_{(A,0)}\mathcal{P}$ be the space of linearised fluctuations $(A,0)\rightarrow(A+\alpha,\,\beta)$ modulo gauge.  
Such fluctuations $(\alpha,\beta)$ around the SD background $(A,0)$ obey the linearised field equations 
\be\label{linYMs}
D_{\gamma}{}^{\dot\gamma}\alpha_{\delta\dot\gamma}=-\frac{\rg^2}{2}\,\beta_{\gamma\delta}\,, \qquad D^{\gamma\dot\gamma}\beta_{\gamma\delta}=0\,,
\ee
where $\beta_{\gamma\delta}=\beta_{(\gamma\delta)}$ are the components of the linearised ASD 2-form: $\beta=\beta_{\gamma\delta}\d x^{\gamma\dot\alpha}\wedge\d x^{\delta}{}_{\dot\alpha}$. The linearised equations induce a splitting of $U$ by a short exact sequence
\be\label{YMsplit}
0\rightarrow U^{+}\hookrightarrow U\rightarrow U^{-}\rightarrow 0\,,
\ee
where $U^+$ is the space of linearised SD solutions modulo gauge:
\be\label{YMSDlin}
 U^{+}=\left\{(\alpha,\beta)\in U\,|\,\beta=0\right\}/\{ \alpha=D\gamma\}\,, \qquad U^-= \{ \beta\in \Omega^{2}_{-}(\mathfrak{g})| D\beta=0\}\,.
\ee
As mentioned in \S\ref{YMTwTh}, the SD field strength of the background means that the space of linearised ASD solutions must be defined asymmetrically, so $U^-$ is given by fluctuations in the ASD curvature, but for which there is \emph{no} canonical SD-free potential. We can, however, define what it means to be purely ASD at a hypersurface using the symplectic form as follows.

For  $\mathscr{A}_{1,2}=(\alpha_{1,2},\,\beta_{1,2}) \in U$
\be\label{YMsymp}
\omega(\mathscr{A}_2 ,\mathscr{A}_{1}):=-\frac{1}{\rg^2}\int_{C}\tr\left(\alpha_{1}\wedge\beta_{2}-\alpha_{2}\wedge\beta_{1}\right)\,,
\ee
It is clear that for $\mathscr{A}_1,\mathscr{A}_2\in U^+$ we have $ \omega(\mathscr{A}_2 ,\mathscr{A}_{1}) =0$ so that $U^+$ is a Lagrangian subspace of $U$. This also gives duality pairing between $(\alpha_1,\beta_1=0)\in U^+$ and $\beta_2\in U^-$ by
\begin{equation}
\int_{C} \tr(\alpha_1\wedge\beta_2)\,, \quad \mbox{ so } \quad U^+=(U^-)^* .\label{pairing}
\end{equation}
This is not hypersurface independent when $\beta_1\neq 0$.  Nevertheless, it can be used to define a splitting of \eqref{YMsplit}, $U=U^+\oplus U^-$, at any given $C$  following~\cite{Mason:2008jy} by:
\begin{defn}\label{DEF:ymasd}
A linearised fluctuation $\mathscr{A}_{1}=(\alpha_1,\beta_1)\in U$ is \emph{ASD at a Cauchy surface} $C$ if
\be\label{ymasympASD}
\int_{C}\tr\left(\alpha_{1}\wedge\beta_{2}\right)=0\,, \qquad \forall\; \beta_2\in U^{-}\,.
\ee
\end{defn}

It is easily checked that at $\scri^+$
\begin{equation}
\int_{\scri^+}\tr\left(\alpha_{1}\wedge\beta_{2}\right)= \int_{\scri^+} \tr\left(\bar\cA^0_1\,\partial_{u}\cA^{0}_{2}\right) \d u\, \D\lambda\, \D\bar\lambda\,.
\end{equation}
So assuming falloff as $u\rightarrow\infty$, the definition for $\mathscr{A}_1$ to be ASD at $\scri^+$ is equivalent to setting $\bar{\cA}^{0}_{1}=0$, as expected.

\medskip

Thus, one can prepare a positive frequency ASD field at $\scri^-$ and measure how much SD part it has acquired from the SD background by the time it gets to $\scri^+$.   This can be measured by pairing with a negative frequency ASD field $\mathscr{A}_2$ at $\scri^+$: this pairing will be the generating functional for the MHV amplitude. So let $\mathscr{A}_{1}$ be ASD at $\scri^-$ and $\mathscr{A}_2$ be ASD at $\scri^+$; evaluating the pairing at $\scri^+$ and using definition~\ref{DEF:ymasd} gives the generating function for the MHV amplitude as 
\be\label{YMgen1}
\begin{split}
\omega(\mathscr{A}_2 ,\mathscr{A}_{1}) &=-\frac{1}{\rg^2}\int_{\scri^+}\tr\left(\alpha_{1}\wedge\beta_{2}\right) \\
 &=-\frac{1}{\rg^2}\int_{\M}\tr\left(D\alpha_{1}\wedge\beta_{2}+\alpha_{1}\wedge D\beta_{2}\right)-\frac{1}{\rg^2}\int_{\scri^-}\tr\left(\alpha_{1}\wedge\beta_{2}\right) \\
 &=\frac{1}{\rg^2}\int_{\M}\tr\left(\beta_1\wedge\beta_2\right)\,,
\end{split}
\ee
upon using Stokes' theorem for $\partial\M=\scri^+ - \scri^-$, the fact that $\mathscr{A}_1$ is ASD at $\scri^-$ and the linearised field equations \eqref{linYMs}. Note that in the final line, we have normalised $\beta_{1,2}\rightarrow \rg^{-2}\beta_{1,2}$ for later convenience.

This $\omega(\mathscr{A}_2 ,\mathscr{A}_{1}) $ is the generating functional for MHV amplitudes, which we wish to expand perturbatively in terms of positive helicity gluons on a fixed SD radiative gauge field background. The right hand side depends implicitly on the background $A$ as the $\beta_{1,2}$ satisfy background-coupled field equations.  The MHV amplitude will be obtained by iteratively perturbing this dependence on $A$.  This would be challenging on space-time, but we will see that on lifting \eqref{YMgen1} to twistor space, where integrability of the SD sector is manifest, it becomes straightforward, leading to a simple extension of the standard MHV amplitude.  

\medskip


\paragraph{Lift to twistor space:} The lift of \eqref{YMgen1} to twistor space is easily achieved using the tools in Section~\ref{YMTwTh}. Observing that the generating functional is equivalent to
\be\label{YMgen2}
\frac{1}{2\,\rg^2}\int_{\M}\d^4 x\,\tr\!\left(\beta_{1}^{\gamma\delta}\,\beta_{2\,\gamma\delta}\right)\,,
\ee
the Penrose transform \eqref{-glpt1} states that $\beta_{1,2}$ are represented in twistor space by cohomology classes $b_{1,2}$ taking values in $H^{0,1}_{\bar{D}}(\PT,\cO(-4)\otimes\End\,E)$, where $E$ and $\bar{D}$ are the Ward bundle and partial connection associated to the SD background, respectively. Consequently, the generating functional is lifted to twistor space using \eqref{GPenrose}:
\be\label{TYMgen1}
 \frac{1}{2\,\rg^2}\int\limits_{\M\times\P^1\times\P^1}\d^{4}x\,\D\lambda_1\,\D\lambda_2\,\la \lambda_1\,\lambda_2\ra^2\,\tr\left[\sH^{-1}_1\,b_1\,\sH_{1}\,\sH^{-1}_{2}\,b_2\,\sH_{2}\right]\,,
\ee
where $\sH$ is the global holomorphic frame for $E|_{X}$. The integral is over two copies of the twistor line $X_{1,2}\cong\P^1\subset\PT$, followed by an integration over the moduli space of lines (i.e., $\M$ itself). The notation $\sH_{1}$ stands for $\sH(x,\lambda_1)$, $b_1$ stands for $b|_{X_1}$ so that $b_1=b_1(x,\lambda_1)$, and so forth.\footnote{The generating functional \eqref{TYMgen1} also follows from a Lagrangian perspective by lifting the fully off-shell Chalmers-Siegel action \eqref{CSaction} to twistor space~\cite{Mason:2005zm} as:
\be\label{YMTA1}
S[a,b]=\frac{\im}{2\,\pi}\int_{\PT}\D^{3}Z\wedge\tr\left(b\wedge F^{(0,2)}[a]\right)
 +\frac{\rg^2}{4}\!\int\limits_{\M\times\P^1\times\P^1}\!\!\!\d^{4}x\,\D\lambda_1\,\D\lambda_2\,\la \lambda_1\,\lambda_2\ra^2\,\tr\left[\sH^{-1}_1\,b_1\,\sH_{1}\,\sH^{-1}_{2}\,b_2\,\sH_{2}\right]\,,
\ee
where $a\in\Omega^{0,1}(\PT,\cO\otimes \End\,E)$ defines $\Dbar=\dbar+a$ on $E\rightarrow\PT$ with $F^{(0,2)}[a]:=[\Dbar,\Dbar]= \dbar a+a\wedge a$, $b\in\Omega^{0,1}(\PT,\cO(-4)\otimes\End\,E)$ and $\D^3 Z$ stands for the canonical section of $\Omega^{3,0}(\PT,\cO(4))$. This twistor action is equivalent to that on space-time; solutions to its equations of motion modulo gauge are in one-to-one correspondence with solutions to \eqref{CSeom}, and the actions agree when evaluated on these corresponding extrema~\cite{Mason:2005zm}. The two-point amplitude of two negative helicity gluon fields evaluated on a SD background is given by the bi-linear portion of the action evaluated on the (on-shell) external states. When $b_1,b_2\in H^{0,1}_{\Dbar}(\PT,\cO(-4)\otimes\End\,E)$ are the two ASD external states with $\Dbar$ on $E$ describing the background, the bi-linear part of the action is then the generating functional \eqref{TYMgen1}. Thus the MHV amplitude follows from background perturbation theory in twistor space itself (cf., \cite{Nair:1988bq,Abe:2004ep,Boels:2007gv}).}


\subsection{Perturbative expansion and MHV amplitude}

We now take the SD background in the generating functional \eqref{TYMgen1} to be constructed from a fixed, classical SD radiative background (treated non-perturbatively) and a finite number of positive helicity gluons propagating on this background. On twistor space, this gives the $\dbar$-operator on $E\rightarrow\PT$ as
\be\label{Pcdecom} 
\dbar+\sa + a:=\Dbar+a\,,
\ee
where $\sa$ is defined by the characteristic data of the SD radiative background via \eqref{radiativeym}, and $a$ is simply a $(0,1)$-form that will later be expanded to finite order as $a=\sum_i a_i$ with each $a_i\in H^{0,1}_{\Dbar}(\PT,\cO\otimes\End\,E)$ encoding SD perturbations/positive helicity gluons on the background. 

In order to expand the generating functional  \eqref{TYMgen1} in $a$ around the background $\Dbar$, note that dependence on $a$ arises only through the holomorphic frame  $\widehat{\sH}$ for the $\dbar$-operator \eqref{Pcdecom}:
\be\label{pertframedef}
\left(\Dbar+a\right)\!|_{X}\widehat{\sH}(x,\lambda)=0\,.
\ee
The kernel of $(\Dbar+a)^{-1}$ acting on sections of $\cO(-1)\otimes E|_{X}$ is given by\footnote{This is the `holomorphic Wilson line' for $\Dbar+a$ along the complex line $X$ of~\cite{Mason:2010yk,Bullimore:2011ni,Adamo:2011pv}.}
\be\label{HWL}
\widehat U_{X}(\lambda,\lambda'):= \frac{1}{2\,\pi\,\im}\,\frac{\widehat{\sH}(x,\lambda)\,\widehat{\sH}^{-1}(x,\lambda')}{\la\lambda\,\lambda'\ra}\,.
\ee
For $\sH$ satisfying $\Dbar|_{X} \sH=0$, write
\be\label{HWL2}
 \widehat U_{X}(\lambda,\lambda')|_{a=0}=\frac{1}{2\,\pi\,\im}\,\frac{\sH(x,\lambda)\,\sH^{-1}(x,\lambda')}{\la\lambda\,\lambda'\ra}\,.
\ee
We can now expand $\widehat{U}_X$ using the geometric series about $a=0$ of $(\Dbar+a)^{-1}$, 
which gives 
\be\label{HWL3}
 \widehat U_{X}(\lambda,\lambda') = \sum_{m=0}^\infty\left(\frac{-1}{2\pi\im}\right)^m\int_{(\P^1)^{m}}\frac{\sH(x,\lambda)}{2\pi\im\,\la \lambda_m\,\lambda'\ra}\left(\prod_{p=1}^m\frac{\sH^{-1}_{p}\,a(\lambda_{p})\,\sH_{p}\,\D\lambda_{p}}{\la \lambda_{p-1}\,\lambda_p\ra}\right)\sH^{-1}(x,\lambda')\,,
\ee
with the identification $\lambda_{0}\equiv\lambda$. 

In terms of $\widehat U_X$, the MHV generating functional is:
\be\label{TYMgen2}
-\frac{2\,\pi^2}{\rg^2}\int\limits_{\M\times\P^1\times\P^1}\d^4x\,\D\lambda_1\,\D\lambda_2\,\la\lambda_1\,\lambda_2\ra^4\,\tr\left(b_1\,\widehat U_{X}(\lambda_1,\lambda_2)\,b_2\,\widehat U_{X}(\lambda_2,\lambda_1)\right)\,,
\ee
so applying \eqref{HWL3} with $a=\sum_{i=1}^{n-2}a_i$, the perturbative expansion to $(n-2)^{\mathrm{th}}$-order in positive helicity gluon wavefunctions is the sum over permutations of the positions of the individual $a_i$ and $b_r$ of 
\begin{multline}\label{TYMgen3}
\rg^{n-2}\int\limits_{\M\times(\P^1)^n}\d^{4}x\,\la\lambda_r\,\lambda_s\ra^{4}\,\tr\left[\sH^{-1}_{1} a_1\sH_{1}\cdots\sH^{-1}_{r-1}a_{r-1}\sH_{r-1}\,\sH^{-1}_{r} b_{r}\sH_{r}\,\sH^{-1}_{r+1}a_{r+1}\sH_{r+1} \right. \\
\left.\cdots  \sH^{-1}_{s-1}a_{s-1}\sH_{s-1}\,\sH^{-1}_{s} b_{s}\sH_{s}\,\sH^{-1}_{s+1}a_{s+1}\sH_{s+1}\cdots \sH^{-1}_{n}a_{n}\sH_{n}\right]\, \prod_{k=1}^{n}\frac{\D\lambda_k}{\la \lambda_k\,\lambda_{k+1}\ra}\,.
\end{multline}
Here, an overall numerical factor has been dropped, the label $n+1\equiv 1$, and the external wavefunctions have been relabeled after performing the expansion so that the negative helicity states are in an arbitrary position in the colour trace.

\medskip

To obtain the momentum space expression for the MHV amplitude, we insert momentum eigenstate representatives \eqref{+glme} and \eqref{-glme} into \eqref{TYMgen3}. As usual, it is possible to be more explicit when the SD radiative background is valued in a Cartan subalgebra; in this case, let each external gluon come with colour vector $\mathsf{T}_{i}$ in some root eigenspace of the Cartan so that $[\sa,\mathsf{T}_i]=e_i\sa$ for some root valued charges $e_i$. The $\P^1$ integrals in \eqref{TYMgen3} are performed against the holomorphic delta functions in the momentum eigenstate representatives, setting $\lambda_i$ to the corresponding $\kappa_i$ momentum spinor. Further, it follows that $\sH_i^{-1}\,a_i\,\sH_i= \e^{e_i g(x,\kappa_i)}a_i$, where $g(x,\kappa_i)$ is the Cartan-valued function defined by the background \eqref{splitym} -- \eqref{Hdef1}. 

Thus, the resulting formula for the colour-ordered MHV amplitude in a Cartan-valued SD radiative background is:   
\be\label{ultimateym}
\rg^{n-2}\frac{\la r\,s\ra^4}{\la1\,2\ra\,\la2\,3\ra\,\cdots\la n-1\;n\ra\,\la n\,1\ra}\int_\M\d^4x\;\exp\left[\sum_{i=1}^n \left(\im\,k_i\cdot x + e_i\,g(x,\kappa_i)\right)\right]\,,
\ee
where charge conservation
\be\label{chargerule}
\sum_{i=1}^n e_i = 0\,,
\ee
must hold in order for the total trace to be non-zero, as the trace is invariant under the action of the Cartan subalgebra. As expected, the SD background does not affect the ratio of ASD (angle bracket) momentum spinor contractions familiar from the Parke-Taylor formula on a flat background. However, 4-momentum conservation is replaced by an integral over Minkowski space which cannot be performed explicitly, since the full functional freedom of the background enters through the function $g$ appearing in the exponential.

The replacement of momentum conservation by such residual space-time integrals is a familiar feature of background perturbation theory; a generic strong background breaks Poincar\'e invariance. However, the expectation from standard background perturbation theory is that an $n$-point tree-level gluon amplitude should have $n-2$ residual space-time integrals in a strong background. It is therefore remarkable that \eqref{ultimateym} contains only a \emph{single} space-time integral regardless of the number of positive helicity gluons in the MHV sector; in this framework, the amplitude behaves like a single contact term.

While surprising from the perspective of the standard Yang-Mills Lagrangian, this is not unexpected from the perspective of the MHV formalism, whereby the Feynman rules of gauge theory are entirely composed of MHV vertices and scalar propagators linking positive and negative helicity legs~\cite{Cachazo:2004kj}. The MHV formalism can be obtained from a particular gauge-fixing of the twistor action \eqref{YMTA1}~\cite{Boels:2007qn,Adamo:2011cb,Adamo:2013cra}, which can be implemented non-perturbatively. The MHV generating functional is equal to the non-local term in this twistor action, which contains only a single space-time integral. Another, less general, perspective is provided by a non-local field redefinition which can be used on space-time to recast the perturbative Yang-Mills Lagrangian in a form where MHV interactions are captured by a single vertex at arbitrary multiplicity~\cite{Mansfield:2005yd,Gorsky:2005sf}, although this is only possible for a subset of SD radiative backgrounds which have commuting null and translational symmetries (such as SDPWs).

\medskip 


\paragraph{Evaluation on SDPWs:} Substituting \eqref{splitsdpwym} into \eqref{ultimateym} enables further simplification of the MHV amplitude on a SDPW background. In this case, the integrand depends only on the lightfront coordinate $x^-$. This allows us to perform the integrals over three of the coordinates -- $x^+,z,\tilde z$ -- to produce a 3-momentum conserving delta function, denoted by $\delta^3_{+,\perp}$ to signify that the $z,\tilde z$ coordinates are transverse to the wave. The remaining lightfront integral is left intact, in keeping with the genericity of the wave-profile $f(x^-)$. The final answer can be written compactly as
\be\label{MHVgluesdpw}
\rg^{n-2}\,\delta^3_{+,\perp}\!\left(\sum_{i=1}^n k_i\right)\,\frac{\la r\,s\ra^4}{\la1\,2\ra\,\la2\,3\ra\,\cdots\la n-1\;n\ra\,\la n\,1\ra}\,\int\limits_{-\infty}^\infty\d x^-\;\exp\,\im\,\F_n(x^-)\,,
\ee
where $\F_n(x^-)$ is the \emph{Volkov exponent}, a common building block of amplitudes in strong electromagnetic fields and laser physics (cf., \cite{Wolkow:1935zz,Seipt:2017ckc}). It is defined by choosing any set of $n-1$ distinct external gluons, say $\{i_{1},\ldots,i_{n-1}\}\subset\{1,\ldots,n\}$, and setting
\be\label{boldK}
\mathbb{K}(x^-) := \sum_{a=1}^{n-1}K_{i_a}(x^-)\,,
\ee
for $K_{i_a}$ the background-dressed null momenta \eqref{sdpwdressed}. Then $\F_n$ is given by
\be\label{Volkovym}
\F_n(x^-) := \int^{x^-}\d s\;\frac{\mathbb{K}^2(s)}{2\,\la\iota|\mathbb{K}(s)|\tilde\iota]} = \sum_{i=1}^n\int^{x^-}\d s\;K_{i\,-}(s)\,,
\ee
with the equivalence of these expressions following from the support of 3-momentum conservation and the fact that $K_i^2=0$ for each external gluon.

This is precisely the MHV amplitude for gluons on a SDPW gauge field background presented in~\cite{Adamo:2020syc}, derived here as a special case of the MHV amplitude on a SD radiative background.


\subsection{Consistency checks}
\label{YMconsist}

It is straightforward to show that the MHV amplitude \eqref{ultimateym} passes basic consistency checks. In appendix~\ref{App:YM}, the formulae are explicitly matched against Feynman diagram computations in background perturbation theory at 3- and 4-points. This becomes unwieldy for higher points $n\geq5$ (indeed, even the $n=4$ calculation is fairly complicated).  Here, we show that the formula has the correct weak background limit.

For the flat background, \eqref{ultimateym} reduces to the standard Parke-Taylor formula for MHV gluon scattering~\cite{Parke:1986gb}: in a trivial background $g(x,\kappa_i)$ becomes independent of $\kappa_i$ so the second term in the exponential of \eqref{ultimateym} vanishes due to charge conservation. The integral over $\M$ can then be performed against the remaining exponential and gives momentum conservation,
\be\label{ymflatlim}
\rg^{n-2}\,\delta^{4}\!\left(\sum_{i=1}^{n}k_i\right)\,\frac{\la r\,s\ra^4}{\la1\,2\ra\,\la2\,3\ra\,\cdots\la n-1\;n\ra\,\la n\,1\ra}\,,
\ee
which is the Parke-Taylor formula.

In the weak field \emph{perturbative limit} of \eqref{ultimateym}, the background is treated to first order and we should recover the Parke-Taylor formula with an additional positive helicity gluon corresponding to the linearised background. Perturbatively, a general SD radiative background is generated by SDPWs, so the perturbative limit corresponds to expanding \eqref{MHVgluesdpw} to first order in $f(x^-)$. This enters only through the Volkov exponent \eqref{Volkovym} to give
\be\label{ympert0}
\delta^{3}_{+,\perp}\!\left(\sum_{i=1}^{n}k_i\right)\,\frac{\la r\,s\ra^4}{\la1\,2\ra\,\la2\,3\ra\,\cdots\la n-1\;n\ra\,\la n\,1\ra}\,\int \d x^{-}\,\sum_{j=1}^{n}\int^{x^{-}}\!\!\d s\,e_{j}\,f(s)\,\frac{\la o\,j\ra}{\la\iota\,j\ra}\,\e^{\im\,k_{j\,-}\,x^{-}}\,.
\ee
To evaluate this perturbative limit, the background is represented by a Fourier mode
\be\label{bFourier}
f(x^-)=\e^{\im\,\omega\,x^{-}}=\e^{\im\,q\cdot x}\,,
\ee
for $q_{\alpha\dot\alpha}:=\omega\,\iota_{\alpha}\tilde{\iota}_{\dot\alpha}$ the on-shell momentum of the background, which can now be viewed as a single positive helicity gluon. Note that if the perturbative background is a pure photon, then $e_j=0$ for all $j=1,\ldots,n$ and \eqref{ympert0} vanishes, as required by the photon-decoupling identity~\cite{Kleiss:1988ne}.

Recalling the definition of the colour-charges, it follows that 
\be\label{pertcc}
e_{j}\,f(s)=\e^{\im\,\omega\,s}\,[\mathsf{t}^{\mathsf{i}},\,\mathsf{T}^{\mathsf{a}_j}]\,,
\ee
for $\mathsf{t}^{\mathsf{i}}$ the generator of the Cartan associated with the background and $\mathsf{T}^{\mathsf{a}_j}$ the generator associated to the external gluon. This means that the sum over $j=1,\ldots,n$ appearing in \eqref{ympert0} is equivalent to summing over the ways of inserting $\mathsf{t}^{\mathsf{i}}$ into the colour-ordering of the original external gluons: 
\begin{multline}\label{ympert1}
\delta^{3}_{+,\perp}\!\left(\sum_{i=1}^{n}k_i\right)\,\frac{\la r\,s\ra^4}{\la1\,2\ra\,\la2\,3\ra\,\cdots\la n-1\;n\ra\,\la n\,1\ra}\,\sum_{j=1}^{n}\left(\frac{\la o\,j\ra}{\la\iota\,j\ra}-\frac{\la o\,j+1\ra}{\la\iota\,j+1\ra}\right) \\
\times\,\frac{-\im}{\omega}\int\d x^{-}\,\exp\left[\im\,\left(\omega+\sum_{i=1}^{n}k_{i\,-}\right)x^{-}\right]\,,
\end{multline}
where we have performed the $\d s$ integral and the sum over $j$ is understood to be cyclic (i.e., $j=n+1=1$). The relative minus sign between the terms in parentheses on the first line arises from the commutator in \eqref{pertcc}.

The $\d x^{-}$ integral in \eqref{ympert1} can now be performed straightforwardly to give
\be\label{ympert2}
\delta^{4}\!\left(q+\sum_{i=1}^{n}\kappa_i\,\tilde{\kappa}_i\right)\,\frac{\la r\,s\ra^4}{\la1\,2\ra\,\la2\,3\ra\,\cdots\la n-1\;n\ra\,\la n\,1\ra}\,\sum_{j=1}^{n}\frac{\la j\,j+1\ra}{\la j\,\iota\ra\,\la\iota\,j+1\ra}\,,
\ee
after using the Schouten identity and dropping an irrelevant overall factor of $-2\pi\im\,\omega^{-1}$. The $j^{\mathrm{th}}$ term in this sum is
\be\label{ympert3}
\delta^{4}\!\left(q+\sum_{i=1}^{n}\kappa_i\,\tilde{\kappa}_i\right)\,\frac{\la r\,s\ra^4}{\la1\,2\ra\,\cdots\la j-1\,j\ra\,\la j\,\iota\ra\,\la\iota\,j+1\ra\,\la j+1\,j+2\ra\cdots\la n-1\;n\ra\,\la n\,1\ra}\,,
\ee
which is the Parke-Taylor formula for $(n+1)$-point MHV scattering in a trivial background, where the additional positive helicity gluon has momentum $q_{\alpha\dot\alpha}$ and is inserted between gluons $j$ and $j+1$ in the colour ordering. This is precisely the expected perturbative limit, with the remaining terms in \eqref{ympert2} giving all other locations of this additional gluon in the colour ordering. 

\medskip

As a final check on the formula, observe that the Volkov exponent in \eqref{ultimateym} is invariant under background gauge transformations. For $A_{\alpha\dot\alpha}(x)\mapsto A_{\alpha\dot\alpha}(x) + \p_{\alpha\dot\alpha}\omega(x)$ for some Cartan-valued $\omega(x)$ on space-time, \eqref{descentym} gives  $g(x,\lambda)\mapsto g(x,\lambda)+\im\,\omega(x)$ so that
\begin{equation*}
\sum_{i=1}^ne_i\,g(x,\kappa_i)\mapsto\sum_{i=1}^ne_i\,g(x,\kappa_i) + \im\,\omega(x)\sum_{i=1}^n e_i = \sum_{i=1}^ne_i\,g(x,\kappa_i)\,,
\end{equation*}
by virtue of the charge conservation \eqref{chargerule}.


\section{N$^k$MHV amplitudes}
\label{SECT:NMHV}

We now consider tree-level amplitudes for gluons and gravitons on self-dual backgrounds in a generic N$^k$MHV helicity configuration. In a trivial background, twistor-string theory~\cite{Witten:2003nn,Berkovits:2004hg,Mason:2007zv,ReidEdwards:2012tq} leads to remarkable formulae for N$^k$MHV amplitudes in gauge theory and gravity as integrals over the moduli of holomorphic maps from the Riemann sphere to twistor space; the degree $d$ of the map is related to the helicity configuration by $d=k+1$~\cite{Roiban:2004yf}. 

As we have seen, the twistor correspondence works just as well with a SD radiative gauge field, so the same principle can be applied to amplitudes on such backgrounds. This again leads to a well-defined formula that is much simpler than might be expected from space-time perturbation theory. There are $4 (k+1)$ twistor moduli integrals at N$^k$MHV. Using space-time Feynman diagrams the number of integrals is given by the space-time dimension multiplied by the number of vertices: $4(n-2)$. Thus, the twistor moduli space counting is as one might expect for a space-time analogue of an MHV vertex expansion in which there are $k+1$ MHV vertices, a significant reduction for general amplitudes when $n>3$. Restricting to a SDPW, there are significant further simplifications arising from the additional symmetries.

In this section, we work with $\cN=4$ super-Yang-Mills (SYM). The inclusion of supersymmetry is just a tool to simplify the resulting formulae; since the expressions are at tree-level, individual particle content is recovered by extracting components from an expansion in the supermomenta. Unlike the MHV amplitudes, these N$^k$MHV formulae are \emph{conjectural}: we do not yet have a systematic way to derive them from first principles. However, we show that they pass several non-trivial consistency checks, such as having the correct flat background and perturbative limits.

\medskip


\paragraph{$\cN=4$ super-Yang-Mills \& the RSVW formula:} Supersymmetric twistor space $\PT$ is an open subset of $\P^{3|4}$ with homogeneous coordinates $Z^{I}=(\mu^{\dot\alpha},\lambda_{\alpha},\chi^{a})$; this is the target space of twistor string theory for $\cN=4$ super-Yang-Mills (SYM). On a trivial background, the tree-level S-matrix of $\cN=4$ SYM is computed by worldsheet correlators in twistor string theory governing holomorphic maps from the Riemann sphere $\P^1$ to $\PT$~\cite{Witten:2003nn,Berkovits:2004hg,Mason:2007zv,ReidEdwards:2012tq}. The resulting Roiban-Spradlin-Volovich-Witten (RSVW) formula~\cite{Roiban:2004yf} for the $\cN=4$ SYM tree-level S-matrix is obtained as an integral over the moduli space of degree $d$ holomorphic maps into twistor space for amplitudes in the N$^{d-1}$MHV helicity sector.

Such a degree $d$ map can be represented as  
\begin{equation}\label{curves}
Z^I(\sigma)=U_{\ba_1\cdots\ba_d}^{I}\,\sigma^{\ba_1}\cdots\sigma^{\ba_d}=:U^{I}_{\ba(d)}\,\sigma^{\ba(d)}\,, 
\end{equation} 
where $\sigma^{\ba}=(\sigma^{\mathbf{0}},\,\sigma^{\mathbf{1}})$ are homogeneous coordinates on the $\P^1$ worldsheet, the parameters $U^{I}_{\ba(d)}$ are the map moduli and $\ba(d)$ is a multi-index denoting a totally symmetric object with $d$ indices. There are $4(d+1)$ bosonic and $4(d+1)$ fermionic degrees of freedom in the moduli. The RSVW formula for the colour-ordered tree-level amplitudes at NMHV degree $d-1$ is then~\cite{Witten:2003nn,Roiban:2004yf}
\begin{equation}
\cA_{n,d}=\int 
\frac{\d^{4|4(d+1)}U}{\mathrm{vol}\,\mathrm{GL}(2,\C)}\,\tr\left(\prod_{i=1}^{n}\frac{a_{i}(Z(\sigma_i)) \,\D\sigma_i}{(i\,i+1)}\right)\,, \label{YMCorrelator}
\end{equation} 
where $(i\,j)=\epsilon^{\mathbf{ab}}\sigma_{i\,\mathbf{b}}\sigma_{j\,\ba}$ is the SL$(2,\C)$-invariant inner product on the homogeneous coordinates of $\P^1$ and $\D\sigma=(\sigma\,\d\sigma)$. Division by the (infinite) volume of GL$(2,\C)$ is understood in the Faddeev-Popov sense, acting on all integrated parameters, and accounts for the $\C^*\times\SL(2,\C)$ degeneracy in the description of the holomorphic map \eqref{curves}. The $a_i\in H^{0,1}(\PT,\cO)\otimes \mathfrak{g}$ are twistor wavefunctions for the external gluon multiplets, usually taken as momentum eigenstates (cf., \cite{Roiban:2004yf,Witten:2004cp,Adamo:2011pv}). The veracity of this formula is established by showing that it has the correct factorization properties~\cite{Vergu:2006np,Skinner:2010cz,Dolan:2011za,Adamo:2013tca}. 

An on-shell $\cN=4$ SYM multiplet is characterised in a momentum eigenstate representation by an on-shell supermomentum $(\kappa_{\alpha}\tilde{\kappa}_{\dot\alpha},\,\kappa_{\alpha}\eta_{a})$, with corresponding twistor representatives (cf., \cite{Adamo:2011pv}): 
\be\label{SYMmeig}
a^{\sa_i}_i(Z)=\sT^{\sa_i}\,\int_{\C^*}\frac{\d s_i}{s_i}\,\bar{\delta}^{2}(\kappa_i-s_i\,\lambda)\,\e^{\im\,s_{i}\,([\mu\,i]+\chi^{a}\,\eta_{i\,a})}\,.
\ee
With such momentum eigenstates inserted in \eqref{YMCorrelator}, the moduli of the $\mu^{\dot\alpha}$ and $\chi^a$ components of the map \eqref{curves} can be integrated out to obtain delta functions~\cite{Roiban:2004yf}:
\begin{multline}
\cA_{n,d}=\int 
\frac{\d^{2(d+1)}\lambda}{\mathrm{vol}\,\mathrm{GL}(2,\C)}\,\delta^{2|4(d+1)}\!\left(\sum_{i=1}^{n}s_i\,(\tilde{\kappa}_i,\eta_i)\,\sigma_{i}^{\ba(d)}\right)\\  \times \prod_{i=1}^{n}\frac{\d s_i\,\D\sigma_i}{s_i\,(i\,i+1)}\,\bar{\delta}^{2}(\kappa_i-s_i\,\lambda(\sigma_i))\,,\label{RSVW}
\end{multline}
with $\lambda_{\alpha}(\sigma)=\lambda_{\alpha\,\ba(d)}\sigma^{\ba(d)}$. The delta functions appearing in this formula are easily seen to imply $4|8$-dimensional supermomentum conservation, and the residual moduli integrals $\{\lambda_{\alpha\ba(d)}, \sigma_i, s_i\}$ -- modulo GL$(2,\C)$ freedom -- are saturated by delta functions to give a residue sum. 

Gluon amplitudes of pure Yang-Mills theory are easily extracted from \eqref{RSVW}: if one is interested in the N$^{d-1}$MHV gluon amplitude where gluons $i_{1},\ldots, i_{d+1}$ are negative helicity, then this is given by the coefficient of $\prod_{k=1}^{d+1}\eta_{i_k}^{4}$ in $\cA_{n,d}$. Recalling the definition of a delta function for Grassmann/anti-commuting variables, it follows that
\be\label{fermomcon}
\delta^{0|4(d+1)}\!\left(\sum_{i=1}^{n}s_i\,,\eta_{i\,a}\,\sigma_{i}^{\ba(d)}\right):=\prod_{\ba(d)}\prod_{a=1}^{4}\left(\sum_{i=1}^{n}s_i\,\eta_{i\,a}\,\sigma_{i}^{\ba(d)}\right)\,,
\ee
so isolating the appropriate coefficient is simply an algebraic exercise. For example, in the case of the MHV amplitude ($d=1$), the required coefficient is
\be\label{fermMHV}
\left.\delta^{0|8}\!\left(\sum_{i=1}^{n}s_i\,,\eta_{i\,a}\,\sigma_{i}^{\ba}\right)\right|_{\eta_{r}^{4}\eta_{s}^{4}}=s_{r}^{4}\,s_{s}^{4}\,\eta_{r}^{4}\,\eta_{s}^{4}\,(r\,s)^{4}\,,
\ee
which leads to the familiar numerator of the Parke-Taylor formula upon integrating out the remaining moduli in \eqref{RSVW}.

\medskip


\paragraph{Amplitudes on a SD radiative background:} The formula \eqref{YMCorrelator} naturally extends to \emph{any} SD background, described on $\PT$ by a holomorphic vector bundle $E\rightarrow \PT$ with partial connection $\Dbar=\dbar+\sa$ obeying $\Dbar^2=0$. On restriction to a rational curve $Z(\sigma)$, this bundle will (generically) be holomorphically trivial, with trivialization $\sH(U,\sigma)$ defined by $\Dbar|_{Z(\sigma)}\sH=0$. Equipped with this holomorphic trivialization, factors of $(i\,i+1)^{-1}$ appearing in \eqref{YMCorrelator} due to Wick contractions in the twistor string worldsheet current algebra are replaced by $\sH_i^{-1} \sH_{i+1}\,(i\,i+1)^{-1}$, where $\sH_i:=\sH(U_r,\sigma_i)$. This leads to a formula for the colour-ordered N$^{d-1}$MHV tree amplitudes on the background: 
\begin{equation}
\cA_{n,d}=\int \frac{\d^{4|4(d+1)}U}{\mathrm{vol}\,\mathrm{GL}(2,\C)}\,\tr\left(\prod_{i=1}^{n}\frac{\sH_i^{-1}\,a_{i}\, \sH_i \,\D\sigma_i}{(i\,i+1)}\right)\,, \label{YMCorr-BG}
\end{equation}
with the external gluon wavefunctions $a_i\in H^{0,1}_{\Dbar}(\PT,\cO\otimes\End\,E)$. For a general, non-abelian background, it is difficult to determine the holomorphic frame $\sH(U,\sigma)$ explicitly, but for a SD radiative background valued in a Cartan $\mathfrak{h}\subset\mathfrak{g}$, $\sH$ can be written as an integral formula.

For such a SD radiative background, there is a higher-degree analogue of the splitting \eqref{splitym}: on restriction to the curve $Z(\sigma)$ there exists some $g(U,\sigma)$ such that $\sa|_{Z(\sigma)}=\dbar|_{Z(\sigma)} g$, with
\begin{equation}
g(U,\sigma)=\frac{1}{2\,\pi\,\im}\int_{\P^1} \frac{\D\sigma'}{(\sigma\,\sigma')}\,\frac{(\sigma\,\xi)}{(\sigma'\,\xi)}\, \sa(Z(\sigma'))\,,\label{g-def}
\end{equation} 
for $\xi\in\P^1$ an arbitrarily chosen point needed to define the inverse of the $\dbar$-operator acting on sections of $\cO$ over $\P^1$. With this, it follows that the holomorphic frame is defined using the exponential map of $\mathfrak{g}$: $\sH_i=\exp[-g(U,\sigma_i)]$. For an external gluon with colour $\mathsf{T}^{\sa_i}$ and charge $e_i$ with respect to the background, it thus follows that
\be\label{hdconj}
\sH^{-1}_{i}\,a_{i}(Z(\sigma_i))\,\sH_{i}=\exp[e_i\,g(U,\sigma_i)]\,a_{i}(Z(\sigma_i))\,.
\ee
With this, the formula for the colour-ordered amplitude on a Cartan-valued SD radiative background is
\begin{equation}\label{YM-SDrad}
\cA_{n,d}=\int\frac{\d^{4|4(d+1)}U}{\mathrm{vol}\,\mathrm{GL}(2,\C)}\,\prod_{i=1}^{n}\e^{e_i\,g(U,\sigma_i)}\,\frac{a_{i}\,\D\sigma_i}{(i\,i+1)}\,.
\end{equation}
Remarkably, all background dependence is captured by a simple exponential factor.

Unlike the RSVW formula \eqref{RSVW}, it is impossible to perform any of the moduli integrals in \eqref{YM-SDrad} on a generic background, even with momentum eigenstates for the external gluon multiplets. This is because $g(U,\sigma)$ introduces generic dependence on the moduli for a general SD radiative background. Nevertheless, the formula \eqref{YM-SDrad} has the remarkable property of posessing fewer integrals than expected from space-time perturbation theory. On a generic SD radiative background, one expects $4(n-2)$ (bosonic) integrals for a $n$-point tree-level gluon amplitude. Evaluated on momentum eigenstates \eqref{SYMmeig} -- now valued in $\End\,E$ -- our formula has $4d$ residual integrals that are not fixed by delta functions; but $d\leq n-3$ for all $n>3$ and thus the background-dressed RSVW formula has fewer residual integrals than space-time perturbation theory in every helicity sector. 

\medskip

If one extracts only external \emph{gluons} from the $\cN=4$ multiplets in \eqref{YM-SDrad}, it is possible to reduce the number of moduli integrations through a judicious reparametrization of the map moduli. Recall that for a N$^{k}$MHV amplitude, $k=d+1$ of the external gluons are negative helicity; without loss of generality, let gluons $r=1,\ldots,d+1$ be negative helicity, with the remainder positive helicity. The parametrization of $Z^{I}(\sigma)$ in terms of the moduli $U^{I}_{\ba(d)}$ can now be traded for
\be\label{Cartp1}
Z^{I}(\sigma)=\sum_{r=1}^{d+1} Z_{r}^{I}\,\prod_{s\neq r}\frac{(\sigma\,s)}{(r\,s)}\,,
\ee
where the $(d+1)$ points in twistor space $\{Z_{r}^{I}\}$ parametrize the new moduli. Note that in this parametrization, the map has the property that $Z^{I}(\sigma_r)=Z^{I}_{r}$.

Substituting this parametrization into \eqref{YM-SDrad}, the fermionic components of the $Z^{I}_{r}$ appear linearly in the exponentials of the negative helicity gluons, and can be integrated out directly to leave only bosonic integrals. This gives the N$^{d-1}$MHV formula as 
\begin{equation}\label{Cart2}
\cA_{n,d}=\int \frac{\mathrm{PT}_{n}}{\mathrm{vol}\,\mathrm{GL}(2,\C)}\,\prod_{r=1}^{d+1}\d^{4}Z_{r}\,\tilde{a}_{r}\,\e^{e_{r}\,g_{r}}\,\prod_{i=d+2}^{n} a_{i}\,\e^{e_{i}\,g_{i}}\,,
\end{equation}
where $\tilde{a}_r$ are the negative helicity gluon wavefunctions, $a_{i}$ are the positive helicity gluon wavefunctions, $g_i:=g(Z_i,\sigma_i)$ and
\be\label{PTfactor}
\mathrm{PT}_{n}:=\prod_{j=1}^{n}\frac{\D\sigma_{j}}{(j\,j+1)}\,,
\ee
with the product corresponding to whatever colour-ordering is desired. Now insert momentum eigenstates 
\begin{equation}
\tilde a_r=\int_{\C^*}\d u_r\,u_r\, \bar\delta^2(\lambda_{r}-u_{r}^{-1}\,\kappa_r)\, \e^{\im\, u_{r}\,[\mu_r\,r]}\, ,\qquad \!\!\!a_i= \int_{\C^*} \frac{\d s_i}{s_i} \bar\delta^2(\kappa_i-s_{i}\lambda(\sigma_i))\, \e^{\im\, s_i\,[\mu(\sigma_i)\,i]}\,,
\end{equation}
with the form of the negative helicity states following from the $Z^I(\sigma_r)=Z^I_r$ property of the parametrization \eqref{Cartp1}.

The moduli integrations in $\d^{2}\lambda_r$ can now be performed against the delta functions in the negative helicity wavefunctions to give:
\begin{multline}\label{Cart3}
\cA_{n,d}=\int\frac{\mathrm{PT}_{n}}{\mathrm{vol}\,\mathrm{GL}(2,\C)}\,\prod_{r=1}^{d+1}\d^{2}\mu_{r}\,\d u_{r}\,u_{r}\,\e^{\im\,u_r\,[\mu_r\,r]+e_{r}\,g_{r}} \\
\times\prod_{i=d+2}^{n} \frac{\d s_{i}}{s_i}\,\bar{\delta}^{2}(\kappa_i-s_i\,\lambda(\sigma_i))\,\e^{\im\,s_i\,[\mu(\sigma_i)\,i]+e_{i}\,g_{i}}\,,
\end{multline}
where 
\be\label{Cart4}
\lambda_{\alpha}(\sigma)=\sum_{r=1}^{d+1}u^{-1}_r\,\kappa_{r\,\alpha} \prod_{s\neq r}\frac{(\sigma\,s)}{(r\,s)}\,.
\ee
Thus, the $g_i$ appearing in the exponents are Cartan-valued functions of $\{\kappa_r,\,\mu_r,\,u_r,\,\sigma_r\}$ as well as $\sigma_i$. In this formula, we obtain $2(n-d-1)$ of the refined/polarized four-dimensional scattering equations\footnote{ The parametrization \eqref{Cart2} is the first step in \cite{Geyer:2016nsh} for proving the formulae of \cite{Geyer:2014fka} from \eqref{YMCorrelator}.} \cite{Geyer:2014fka,Geyer:2016nsh}, written in this parametrization as  
\begin{equation}\label{refscatt}
s_i\,\lambda_{\alpha}(\sigma_i)-\kappa_{i\,\alpha}=s_i\,\sum_{r=1}^{d+1} u^{-1}_r\,\kappa_{r\,\alpha}\,\prod_{s\neq r}\frac{(i\,s)}{(r\,s)}-\kappa_{i\,\alpha}=0\, .
\end{equation}
The remaining $2(d+1)$ refined scattering equations would have arisen from the $\d^{2}\mu_r$ integrals, but these cannot be performed analytically due to the non-trivial background. 

In \eqref{Cart3} there are $2(n+d+1)$ integrations against $2(n-d-1)$; taking into account the GL$(2,\C)$ quotient this leaves $4d$ free integrations. Of course, this matches the number of residual integrals in the general formula \eqref{YMCorr-BG}, but this parametrization enables us to explicitly remove $2|4(d+1)$ of the integrals against delta functions. A further $2(n-d-1)$ integrals can be explicitly removed against the remaining delta functions in \eqref{Cart3} by solving \eqref{refscatt} for $\sigma_i$ in terms of the $\{u_r, \sigma_r,\kappa_r, \kappa_i\}$, but the resulting formulae are not so enlightening.  Note that the same computations can be performed starting from \eqref{YMCorr-BG} without assuming that $\sa$ lies in a Cartan subalgebra, as momentum eigenstates are given by the expression \eqref{SYMmeig} on any SD radiative background.

\medskip


\paragraph{Evaluation on SDPWs:} The formula \eqref{YM-SDrad} can be further simplified when the background is restricted to be a SDPW. In this case, recall that the background partial connection pulled back to the curve $Z(\sigma)$ is
\be\label{twistorsdpwym1}
\begin{split}
\sa(Z(\sigma)) &= 2\pi\,\int_{\C^*}\frac{\d t}{t}\,\bar\delta^2(\iota-t\,\lambda(\sigma))\,\F(t[\mu(\sigma)\,\tilde\iota]) \\
 &=2\pi\,\la o\,\lambda(\sigma)\ra\,\bar{\delta}(\la\iota\,\lambda(\sigma)\ra)\,\F\left(\frac{[\tilde\iota\,\mu(\sigma)]}{\la o\,\lambda(\sigma)\ra}\right)\,.
\end{split}
\ee
From this, we see that the only contributions to \eqref{g-def} arise when $\la\iota\,\lambda(\sigma)\ra=0$, so it makes sense to parametrize the moduli of the map $\lambda_{\alpha}(\sigma)$ in a way that makes this manifest. Now, $\la\iota\,\lambda(\sigma)\ra$ is a homogeneous function of degree $d$ on $\P^1$; in an affine patch $\sigma^{\alpha}=(1,z)$ this means that $\la\iota\,\lambda(z)\ra$ is a degree $d$ polynomial in $z$. Let $\{b_1,\ldots,b_d\}$ denote the $d$ roots of this polynomial, and $b_0$ encode the overall scale so that:
\be\label{ilamb1}
\la\iota\,\lambda(z)\ra=b_{0}\,\prod_{r=1}^{d}(z-b_{r})\,.
\ee
These $\{b_0,b_1,\ldots,b_d\}$ provide an alternative parametrization of the moduli of $\la\iota\,\lambda(\sigma)\ra$; written in homogeneous coordinates this parametrization is simply
\be\label{ilamb2}
\la\iota\,\lambda(\sigma)\ra=\la\iota\,b_0\ra\,\prod_{r=1}^{d}(\sigma\,b_r)\,,
\ee
with each $b_r\in\P^1$ constituting only a single degree of freedom in the moduli measure. 

This parametrization also induces a natural choice of basis for $H^0(\P^1,\cO(d))$:
\be\label{pbasis}
\mathfrak{s}_{0}(\sigma)=\frac{\la\iota\,\lambda(\sigma)\ra}{\la\iota\,\lambda(b_0)\ra}\,, \qquad \mathfrak{s}_{r}(\sigma)=\frac{(\sigma\,b_0)}{(b_{r}\,b_0)}\,\prod_{s\neq0, r}\frac{(\sigma\,b_{s})}{(b_{r}\,b_{s})}\,, \quad r=1,\ldots,d\,.
\ee
The map components $\la o\,\lambda(\sigma)\ra$ can now be expanded in this basis as
\be\label{olamb1}
\la o\,\lambda(\sigma)\ra=\nu_{0}\,\mathfrak{s}_{0}(\sigma)+\sum_{r=1}^{d}\nu_{r}\,\mathfrak{s}_{r}(\sigma)\,,
\ee
with $\{\nu_0,\nu_1,\ldots,\nu_d\}$ the new parametrization of the map moduli. This parametrization has the nice property that
\be\label{olamb2}
\la o\,\lambda(b_k)\ra=\nu_k \quad \forall\, k=0,1,\ldots,d\,.
\ee
Finally, the $\mu^{\dot\alpha}(\sigma)$ components of \eqref{curves} can be decomposed as
\be\label{mucomp}
\mu^{\dot\alpha}(\sigma)=\tilde{\iota}^{\dot\alpha}\,\tilde{m}_{\ba(d)}\,\sigma^{\ba(d)}+\tilde{o}^{\dot\alpha}\,m_{\ba(d)}\,\sigma^{\ba(d)}\,,
\ee
with $\{\tilde{m}_{\ba(d)},m_{\ba(d)}\}$ the $2d+2$ moduli of the map. 

Feeding these parametrizations of the map into \eqref{g-def}, one obtains
\be\label{g-def1}
g(U,\sigma)=-\frac{\im}{b_0}\sum_{r=1}^{d}\frac{\nu_{r}\,(\sigma\,\xi)}{(\sigma\,b_r)\,(b_r\,\xi)}\,\F\!\left(\frac{m_{\ba(d)}\,b_{r}^{\ba(d)}}{\nu_{r}}\right)\,\prod_{s\neq0, r}\frac{1}{(b_{r}\,b_{s})}\,.
\ee
The $d+1$ moduli $m_{\ba(d)}$ of $[\tilde{\iota}\,\mu(\sigma)]$ can be conveniently reparametrized by  
\be\label{xdef}
y := \frac{m_{\ba(d)}\,b_{0}^{\ba(d)}}{\la\iota\,\lambda(b_0)\ra}\,,\qquad x_{r}:=\frac{m_{\ba(d)}\,b_{r}^{\ba(d)}}{\nu_{r}}\,,\quad r=1,\dots,d\,,
\ee
so that
\be\label{mucomp2}
[\tilde{\iota}\,\mu(\sigma)]=y\,\la\iota\,\lambda(\sigma)\ra+\sum_{r=1}^{d}x_{r}\,\nu_{r}\,\mathfrak{s}_{r}(\sigma)\,.
\ee
This parametrization makes it clear that $g$ depends on only the $\{x_r\}$ moduli of the $\mu^{\dot\alpha}$ map:
\be\label{g-def2}
g(x,\sigma)=-\frac{\im}{b_0}\sum_{r=1}^{d}\frac{\nu_{r}\,(\sigma\,\xi)}{(\sigma\,b_r)\,(b_r\,\xi)}\,\F(x_r)\,\prod_{s\neq0, r}\frac{1}{(b_{r}\,b_{s})}\,,
\ee
with dependence on the moduli of the $\lambda_{\alpha}$ map implicit on the left-hand side.\footnote{For instance, the expression \eqref{splitsdpwym} for $g$ at degree $d=1$ is obtained by choosing the reference spinor to be $\xi=o$ and identifying $\lambda$ with $\sigma$ by using the GL$(2,\C)$ symmetry. In this case, $x_1\equiv x^-$.}

Inserting \eqref{g-def2} into the general formula and evaluating on momentum eigenstates, the moduli integrals in $\tilde{m}_{\ba(d)}$ and $y$ as well as all of the fermionic moduli for $\chi^{a}$ can be performed to yield delta functions. The result is:
\begin{multline}\label{YMamp-pw-fin}
\cA_{n,d}=\delta\!\left(\sum_{i=1}^{n}[\tilde{o}\,i]\,\la\iota\,i\ra\right)\,\int\frac{\d^{d+1}b\,\d^{d+1}\nu\,\d^{d}x}{\mathrm{vol}\,\mathrm{GL}(2,\C)}\,J(b,\nu)\,\delta^{1|4(d+1)}\!\left(\sum_{j=1}^{n}s_{j}\,([\tilde{\iota}\,j],\eta_j)\,\sigma_{j}^{\ba(d)}\right) \\
\times \prod_{i=1}^{n}\frac{\d s_i\,\D\sigma_i}{s_i\,(i\,i+1)}\,\bar{\delta}^{2}(\kappa_i-s_{i}\,\lambda(\sigma_i))\,\e^{\im\,\varphi_{i}}\,,
\end{multline}
where $J(b,\nu)$ is the Jacobian resulting from the various reparametrizations performed on the map moduli:
\be\label{Jacobian}
J(b,\nu)=\frac{b_{0}^{d+1}\,\nu_1\cdots \nu_{d}}{|b_0\,b_{1}\cdots b_{d}|}\,,
\ee
for $|b_0\,b_{1}\cdots b_{d}|$ the Vandermonde determinant, and
\be\label{dVolkovYM}
\varphi_{i}:=s_{i}\,\sum_{r=1}^{d} x_{r}\,\nu_{r}\,[\tilde{o}\,i]\,\mathfrak{s}_{r}(\sigma_i)+e_i\,g(x,\sigma_i)\,,
\ee
is the generalized Volkov exponential associated with degree-$d$ maps into twistor space.

Compared to \eqref{YM-SDrad} for a general SD radiative background, there are fewer residual moduli integrals in a SDPW. It is easy to see that the collection of delta functions in \eqref{YMamp-pw-fin} imply momentum conservation in the $x^+$, $z$ and $\tilde{z}$-directions, as expected for scattering amplitudes in a plane wave background. Taking into account this 3-momentum conservation and the GL$(2,\C)$ quotient, there are $2d-1$ residual integrals in $\cA_{n,d}$ that are \emph{not} saturated by delta functions. This counting matches the number of integrations expected for a N$^{d-1}$MHV amplitude computed with MHV diagrams in a SDPW background: one integral for each MHV vertex and one integral for each scalar propagator. 

\medskip

While \eqref{YM-SDrad} -- and consequently \eqref{YMamp-pw-fin} -- are conjectural, it is easy to see that these formulae pass several non-trivial tests of their validity. Indeed, background gauge invariance and the correct trivial background and perturbative limits follow immediately by straightforward generalisations of the arguments of section~\ref{YMconsist}. Furthermore, when $d=1$ the formulae are equal to the MHV expressions \eqref{ultimateym} and \eqref{MHVgluesdpw}. In the latter case, this follows after using the GL$(2,\C)$ quotient to fix $b_{0}=1$, $b^{\alpha}_{1}=\iota^{\alpha}$, $\nu_0=0$ and $\nu_1=-1$, and setting $\xi_{\alpha}=o_{\alpha}$ in the $d=1$ version of \eqref{YMamp-pw-fin}.

Finally, we re-emphasize that $\cN=4$ supersymmetry is playing a trivial role here: pure gluon amplitudes are extracted from these formulae algebraically, in exactly the same way as in a flat background, by expanding the fermionic delta functions as described in \eqref{fermomcon}.


\section{Further discussion}\label{discussion}

In this paper, we proved the gauge theory formulae on a strong self-dual plane wave background announced in~\cite{Adamo:2020syc} from first principles and extended it to an arbitrary self-dual radiative background. In addition, we gave conjectural formulae for the \emph{full} tree-level gluon S-matrix on any self-dual radiative background; these require many fewer integrations than their corresponding counterparts computed via standard space-time Feynman diagrams in background perturbation theory. An explanation for why our formulae contain so many fewer integrations than what is na\"ively expected is provided by a background-dressed version of the \emph{MHV formalism}. In a trivial background, the MHV formalism is an alternative set of Feynman rules for perturbative Yang-Mills theory in which the only vertices are MHV vertices, where legs of opposite helicity are connected by a scalar propagator that includes an off-shell prescription for the un-dotted momentum spinors~\cite{Cachazo:2004kj,Risager:2005vk}.

Applying background perturbation theory to the twistor action \eqref{YMTA1} of Yang-Mills theory defines the MHV vertices for such a background-dressed MHV formalism, and fixing an appropriate axial gauge (to eliminate the cubic $\overline{\mbox{MHV}}$ vertex) makes these the \emph{only} vertices in the theory. In the special case of a SDPW background, this can be done explicitly and a formula for the propagator on twistor space can be found (see appendix \ref{propapp}):
\begin{multline}\label{mhvprop}
\scG^{\msf{ab}}(x_1,\lambda_1;\,x_2,\lambda_2) =  \delta^{\msf{a}\msf{b}}\int\frac{\d^4k}{(2\pi)^4}\left[\frac{1}{k^2}\,\bar\delta^1_0(\la\lambda_1|k|\tilde\iota])\wedge\bar\delta^1_{-4}(\la\lambda_2|k|\tilde\iota])\right.\\
\left.+ \frac{1}{4\pi}\left(\frac{\la\hat\lambda_1|\d x_1|\tilde\iota]\wedge\bar\delta^1_{-4}(\la\lambda_2\,\lambda_1\ra)}{\la\lambda_1\,\hat\lambda_1\ra\,\la\lambda_1|k|\tilde\iota]} + \frac{\la\hat\lambda_2|\d x_2|\tilde\iota]\wedge\bar\delta^1_{0}(\la\lambda_1\,\lambda_2\ra)}{\la\lambda_2\,\hat\lambda_2\ra\,\la\lambda_2|k|\tilde\iota]}\right)\right]\e^{\im(\phi(x_1)-\phi(x_2))}\,,
\end{multline}
where $\phi(x)$ is given by \eqref{sdpwHJ}, having taken $k$ off-shell, and the various delta functions are defined by
\be\label{delta1m}
\bar\delta^1_m(\la\lambda\,\kappa\ra) = \int_{\C^*}\frac{\d s}{s^{m+1}}\,\bar\delta^2(\kappa-s\,\lambda)\,.
\ee
This is structurally equivalent to the trivial background propagator, with the axial gauge defined by the spinor $\tilde\iota_{\dot\alpha}$ associated with the SDPW \cite{Adamo:2011cb, Boels:2007qn}, but with the extra phase factor $\e^{\im(\phi(x_1)-\phi(x_2))}$. It would be interesting to develop the MHV formalism on \emph{any} SD radiative background, and to compare the formulae arising from such a `disconnected' prescription to the `connected' formulae presented here; Green's functions on more general self-dual backgrounds were already obtained explicitly in \cite{Atiyah:1981ey}, albeit not in MHV gauge. If an exact matching were possible, then this would be a route towards proving our formulae; however, even on a trivial background the connected and disconnected prescriptions have only been partially linked (cf., \cite{Gukov:2004ei}) -- even through each is known to be correct. 

More generally, the usual unitarity-based arguments which prove the original RSVW formula cannot be directly applied to scattering amplitudes on a SD radiative background, as the functional freedom of the background means that even tree-level amplitudes are not rational functions of the kinematic data. Of course, unitarity and its consequences must still leave their imprint on the structure of amplitudes in a background, but it is not immediately clear how these can be operationalized in a way to constrain or prove whether a formula for a given scattering amplitude is correct. Nevertheless, it was recently shown that the analytic structure of 4-point amplitudes in strong field QED can be constrained by gauge invariance~\cite{Ilderton:2020rgk}; extending these findings to non-abelian gauge theory could provide a powerful new tool to check our formulae for N$^{k\geq1}$MHV amplitudes on a SD radiative background.

One restriction in our work was to radiative backgrounds.  This excludes charged background fields or instantons.  The main reason for this was technical in the sense that such solutions have nontrivial topology.  Nevertheless, the underlying twistor theory is well-developed for such solutions so that it wouldn't be too hard to extend our methods here. The application to amplitudes on instanton backgrounds was already addressed in  \cite{Jiang:2008xw}.  Charged backgrounds might provide a starting point for applications to gravitational waves.

An obvious goal for future research is to find all-multiplicity expressions for the tree-level S-matrix of Yang-Mills theory on \emph{any} radiative background, rather than the chiral cases considered here. An obvious route towards this is \emph{ambitwistor string theory}~\cite{Mason:2013sva}, which is a non-chiral formalism that produces compact, all-multiplicity formulae for gauge theory scattering at tree-level and beyond on a trivial background~\cite{Cachazo:2013hca}. Ambitwistor string theory can be coupled to a fixed background gauge field and is anomaly free (at genus zero) precisely when the background obeys the Yang-Mills equations~\cite{Adamo:2018hzd}; furthermore, the fixed vertex operators for gluon perturbations on any gauge field background are known~\cite{Adamo:2018ege}, and have been used to explicitly calculate 3-point amplitudes on a (real, non-chiral) plane wave in any number of space-time dimensions~\cite{Adamo:2017sze}. To extend these computations beyond 3-points an understanding of integrated vertex operators (or equivalently, the scattering equations) in a background gauge field is required. 

Nevertheless, we observe that there are potential phenomenological applications of our formulae -- even though the backgrounds are chiral. In particular, back-reaction effects from probes on strong gauge fields (cf., \cite{Epelbaum:2013waa,Seipt:2016fyu}) can be studied by considering scattering on a complexified background gauge field~\cite{Ilderton:2017xbj}. It would be very interesting to explore if our results can be of use in providing high-multiplicity data in the study of back-reaction in strong field QED or QCD.

Finally, in~\cite{Adamo:2020syc} we also presented all-multiplicity formulae for MHV scattering of gravitons in a strong self-dual gravitational plane wave. The proof and generalizations of those formulae will appear in a separate upcoming publication; the theory is quite distinct from the gauge theory case studied here.

\acknowledgments

TA is supported by a Royal Society University Research Fellowship. AS is supported by a Mathematical Institute Studentship, Oxford. LJM is partially supported by STFC grant ST/T000864/1.


\appendix

\section{3- and 4-particle checks on SD plane waves}
\label{App:YM}

In this appendix, we compute the 3- and 4-particle gluon and graviton MHV amplitudes in SDPWs directly from Feynman diagrams in background perturbation theory. This serves as a check on the MHV formulae derived in sections \ref{SECT:MHVgauge}, which are shown to match the Feynman diagram computations for $n=3,4$.

The Feynman rules for perturbative Yang-Mills theory on a general gauge theory plane wave background (and in generic space-time dimension) can be found in~\cite{Adamo:2017nia, Adamo:2018mpq}, and are easily specialized to the case of a SDPW background. Vertices are simply read off from the background field Lagrangian, while the gluon propagator $\scG^{\msf{ab}}_{ab}(x,y)$ obeys
\be\label{gluonpropeqn}
\left(D^2\,\delta^b_a +2\,\im\,e\,F_a{}^b\right)\scG^{\mathsf{ab}}_{bc}(x,y) =\delta^{\mathsf{ab}}\, \eta_{ac}\,\delta^4(x-y)\,,
\ee
where $D_{a}=\partial_{a}-\im\,A_{a}$ is the background covariant derivative, $A_{a}$ is the SDPW gauge field, $F_{ab}$ is its field strength, and $e$ is the charge of the propagated gluon with respect to the Cartan-valued background. The Green's function solving this equation is given by
\be\label{gluonprop}
\scG^{\mathsf{ab}}_{ab}(x,y)= \frac{\cN\,\delta^{\msf{a}\msf{b}}}{2\pi\im}\int\frac{\d^4k}{k^2+\im\,\varepsilon}\,\D_{ab}(x^-,y^-)\,\e^{\im(\phi(x)-\phi(y))}\,,
\ee
where $\cN = -\pi\im/(2\pi)^4$ is a convenient normalisation factor. The phase factors $\phi(x)$ are defined by \eqref{sdpwHJ}, now understood to be evaluated at the off-shell momentum $k$. The tensor structure $\D_{ab}(x^-,y^-)$ is
\be\label{gluontensor}
\D_{ab}(x^-,y^-) = \epsilon_{\alpha\beta}\left(\epsilon_{\dot\alpha\dot\beta}+\frac{e\,\Delta f}{k_+}\,\tilde{\iota}_{\dot\alpha}\,\tilde{\iota}_{\dot\beta}\right)\,,
\ee
where $\Delta f:=f(x^-)-f(y^-)$.

\medskip

The tree-level 3-point gluon amplitude is read off from the cubic interaction in the background field Lagrangian~\cite{Adamo:2017nia}:
\be\label{3ptstefan}
\cA_3 = 2\,\im\,\rg\,f^{\msf{abc}}\,\delta^3_{+,\perp}\!\left(\sum_{i=1}^3k_i\right)\int\d x^-\left(\cE_1\cdot\cE_3\,K_1\cdot\cE_2 + \text{cyclic}\right)\,\e^{\im\,\F_3(x^-)}\,,
\ee
where $\cE_{i}$ are the background-dressed polarizations of the external gluons, $K_{i}$ are the dressed momenta and $\F_3(x^-)$ is the 3-point Volkov exponent of \eqref{Volkovym}. It is straightforward to see that $\cA_3$ vanishes when all three external gluons have the same helicity; for the MHV configuration we take gluons 1 and 2 to be negative helicity and gluon 3 to be positive helicity. Making a judicious choice for the reference spinor $\xi_{\alpha}$ in the positive helicity polarization \eqref{Aplus} (i.e., a judicious gauge choice) substantially simplifies the calculation; we use $\xi_{\alpha}=\kappa_{2\,\alpha}$ so that
\be\label{3ptpolym}
\cE_{i\,\alpha\dot\alpha} = \frac{\kappa_{i\,\alpha}\,\tilde\iota_{\dot\alpha}}{[\tilde\iota\,i]}\,,\quad i=1,2\,;\qquad\cE_{3\,\alpha\dot\alpha} = \frac{\kappa_{2\,\alpha}\,\tilde K_{3\,\dot\alpha}}{\la2\,3\ra}\,.
\ee
In this case, $\cE_1\cdot\cE_2 = \cE_2\cdot\cE_3 = 0$, and the integrand in $\cA_3$ reduces to
\begin{equation*}
\cE_1\cdot\cE_3\,K_1\cdot\cE_2 + \text{cyclic} = -\frac{\la1\,2\ra^2}{\la2\,3\ra}\,\frac{[\tilde\iota\,3]}{[\tilde\iota\,2]}+\text{cyclic}\,.
\end{equation*}
Now, in general, 3-momentum conservation in the $x^+,z,\tilde z$-directions can be written in the equivalent forms
\be\label{3momcon}
\begin{split}
\sum_{i=1}^n \la o\,i\ra\,[\tilde\iota\,i] &= \sum_{i=1}^n \la \iota\,i\ra\,[\tilde o\,i] = \sum_{i=1}^n \la\iota\,i\ra\,[\tilde\iota\,i] = 0\\
\Leftrightarrow \sum_{i=1}^n\la\iota\,i\ra\,\tilde K_{i\,\dot\alpha} &= \sum_{i=1}^n\la\iota\,i\ra\,\tilde\kappa_{i\,\dot\alpha} = 0 = \sum_{i=1}^n\kappa_{i\,\alpha}\,[\tilde\iota\,i]\,,
\end{split}
\ee
using charge conservation $\sum_{i=1}^ne_i=0$. For $n=3$ this implies that $\la 1\,3\ra\,[\tilde\iota\,3] = -\la 1\,2\ra\,[\tilde\iota\,2]$, which in turn simplifies the 3-point MHV amplitude to
\be\label{3ptmhvym}
\cA_3^\text{MHV} = -2\,\im\,\rg\,f^{\msf{abc}}\,\delta^3_{+,\perp}\!\left(\sum_{i=1}^3k_i\right)\frac{\la1\,2\ra^3}{\la2\,3\ra\,\la3\,1\ra}\int\d x^-\,\e^{\im\,\F_3}\,.
\ee
which matches the formula \eqref{MHVgluesdpw} at $n=3$.

For completeness, we note that the 3-point $\overline{\text{MHV}}$ is also easily computed from the background field Lagrangian, and is again simplified on the support of 3-momentum conservation:
\be\label{3ptmhvbarym}
\cA_3^{\overline{\text{MHV}}} = -2\,\im\,\rg\,f^{\msf{abc}}\,\delta^3_{+,\perp}\!\left(\sum_{i=1}^3k_i\right)\int\d x^-\,\e^{\im\,\F_3}\,\frac{[\![1\,2]\!]^3}{[\![2\,3]\!]\,[\![3\,1]\!]}(x^-)\,,
\ee
where $[\![i\,j]\!](x^-):=\tilde K_i^{\dot\alpha}(x^-)\,\tilde K_{j\,\dot\beta}(x^-)$, and gluons 1 and 2 are positive helicity. Note that -- in contrast to amplitudes on a trivial background -- $\cA_3^{\overline{\text{MHV}}}$ is \emph{not} the parity conjugate of $\cA_3^\text{MHV}$, as the former contains background dressed spinor contractions. This is a result of the chirality of the SDPW background itself.

\medskip

The tree-level 4-gluon amplitude on a SDPW background recieves contributions from four Feynman diagrams~\cite{Adamo:2018mpq}:
\be\label{4ptdecomym}
\cA_4 = \cA_\msf{s} + \cA_\msf{t} + \cA_\msf{u} + \cA_\text{cont}\,,
\ee
where the first three terms represent the $\msf{s}$, $\msf{t}$ and $\msf{u}$ channels, and the last term is the contact diagram coming from the quartic interaction. The simplest of these is the contact contribution:
\be\label{contym}
\begin{split}
\cA_\text{cont} = \frac{\rg^2}{\cN}\,\delta^3_{+,\perp}\!\left(\sum_{i=1}^4k_i\right)\int\d x^-\,&\e^{\im\F_4}\,\Bigl[f^{\msf{a}_1\msf{a}_2\msf{b}}\,f^{\msf{a}_3\msf{a}_4\msf{b}}\left(\cE_1\cdot\cE_3\,\cE_2\cdot\cE_4-\cE_1\cdot\cE_4\,\cE_2\cdot\cE_3\right)\\
&+ f^{\msf{a}_1\msf{a}_3\msf{b}}\,f^{\msf{a}_2\msf{a}_4\msf{b}}\left(\cE_1\cdot\cE_2\,\cE_3\cdot\cE_4-\cE_1\cdot\cE_4\,\cE_2\cdot\cE_3\right) \\
&+ f^{\msf{a}_1\msf{a}_4\msf{b}}\,f^{\msf{a}_2\msf{a}_3\msf{b}}\left(\cE_1\cdot\cE_2\,\cE_3\cdot\cE_4 - \cE_1\cdot\cE_3\,\cE_2\cdot\cE_4\right)\Bigr]\,.
\end{split}
\ee
The exchange diagrams contain a propagator insertion; for example, the $\msf{s}$-channel is given by
\be\label{schannelym}
\begin{split}
\cA_\msf{s} = &\;\rg^2\,\delta^3_{+,\perp}\!\left(\sum_{i=1}^4k_i\right)f^{\msf{a}_1\msf{a}_2\msf{b}}\,f^{\msf{a}_3\msf{a}_4\msf{b}}\int\d^2\mu[\msf{s}]\,\D^\msf{s}_{ab}(x^-,y^-)\\
&\times\bigl[\cE_1\cdot\cE_2\,(K_1-K_2)^a + 2\,\cE_1\cdot K_2\,\cE_2^a-2\,\cE_2\cdot K_1\,\cE_1^a\bigr](x^-)\\
&\times\bigl[\cE_3\cdot\cE_4\,(K_4-K_3)^b - 2\,\cE_3\cdot K_4\,\cE_4^b+2\,\cE_4\cdot K_3\,\cE_3^b\bigr](y^-) + (x^-\leftrightarrow y^-)\,,
\end{split}
\ee
where the propagator tensor structure $\D_{ab}^\msf{s}$ is defined with respect to exchanged momentum $k_1+k_2$ and charge $e_1+e_2$. The integral is over two lightfront coordinates $x^-$ and $y^-$, with the integration measure
\be\label{d2mus}
\begin{split}
&\d^2\mu[\msf{s}] = \Theta(x^--y^-)\,\frac{\d x^-\;\d y^-}{(k_1+k_2)_+}\,\exp\!\left[\im\sum_{i=1,2}\int^{x^-}\!\!K_{i\,-}(s)\,\d s\right.\\
 &\left.+ \im\sum_{j=3,4}\int^{y^-}\!\!K_{j\,-}(s)\,\d s - \frac{\im\,(k_1+k_2)}{\,(k_1+k_2)_+}\int_{y^-}^{x^-}\!\left(\tilde k_1+\tilde k_2 + (e_1+e_2)\,f(s)\right)\d s\right].
\end{split}
\ee
There are similar formulae for the $\mathsf{t}$- and $\mathsf{u}$-channels~\cite{Adamo:2018mpq}.

\emph{A priori}, these expressions look a long way off our MHV formula \eqref{MHVgluesdpw}, which contains only a single lightfront integral at arbitrary multiplicity. However, intricate cancellations in the integrands enable the application of some non-trivial integration-by-parts (IBP) identities (cf., \cite{Adamo:2018mpq}) which reduce all space-time expressions to a single lightfront integral. Let particles $1,2$ be negative helicity and $3,4$ be positive helicity, with dressed polarizations given by
\be\label{4ptpolym}
\cE_{i\,\alpha\dot\alpha} = \frac{\kappa_{i\,\alpha}\,\tilde\iota_{\dot\alpha}}{[\tilde\iota\,i]}\,,\quad i=1,2\,;\qquad\cE_{j\,\alpha\dot\alpha} = \frac{\kappa_{1\,\alpha}\,\tilde K_{j\,\dot\alpha}}{\la1\,j\ra}\,,\quad j=3,4\,.
\ee
This gauge choice eliminates the contact contribution to the 4-point amplitude completely. The $\msf{s}$-channel diagram evaluates to
\be\label{smhvym}
\begin{split}
\cA_\msf{s}^\text{MHV} = 4\,\rg^2\,\delta^3_{+,\perp}\!\left(\sum_{i=1}^4k_i\right)&f^{\msf{a}_1\msf{a}_2\msf{b}}f^{\msf{a}_3\msf{a}_4\msf{b}}\frac{\la1\,2\ra^3}{\la4\,1\ra}\frac{[\tilde\iota\,2]}{[\tilde\iota\,1]\la1\,3\ra}\\
&\times\int\d^2\mu[\msf{s}]\,[\![3\,4]\!](y^-) + (x^-\leftrightarrow y^-)\,.
\end{split}
\ee
To reduce this to a single lightfront integral, consider the following manipulation:
\be\label{newidstepym}
\begin{split}
\frac{1}{\cN}\int&\d^4 x\;\exp\!\left[\im\sum_{i=1}^4\phi_i(x)\right]\\
&= \frac{1}{\cN}\int\d^4 x\,\d^4y\;\delta^4(x-y)\,\exp\!\left[\im\sum_{i=1,2}\phi_i(x) + \im\sum_{j=3,4}\phi_j(y)\right].
\end{split}
\ee
Now, let $\scG^{\mathsf{ab}}(x,y)$ be the propagator for an adjoint-valued scalar field of charge $e$ with respect to the background:
\be\label{chargedscprop}
\scG^{\mathsf{ab}}(x,y) = \frac{\cN\,\delta^{\msf{a}\msf{b}}}{2\pi\im}\,\int\frac{\d^4k}{k^2+\im\,\varepsilon}\,\e^{\im(\phi(x)-\phi(y))}\,,
\ee
which obeys
\be\label{chargedKG}
D^2\scG^{\mathsf{ab}}(x,y) = \delta^{\mathsf{ab}}\,\delta^4(x-y)\,.
\ee
This identity can be used to replace the $\delta^{4}(x-y)$ in \eqref{newidstepym}, following which we integrate the charged Laplacian $D^2$ by parts.

The crux of the resulting IBP identity is the relation
\be\label{ibpym}
\frac{1}{\cN}\int\d^4 x\;\exp\!\left[\im\sum_{i=1}^4\phi_i(x)\right] =  -\delta^3_{+,\perp}\!\left(\sum_{i=1}^4k_i\right)\!\int\d^2\mu[\msf{s}]\,K_3\cdot K_4(y^-) + (x^-\leftrightarrow y^-)\,.
\ee
On the right hand side, $K_3\cdot K_4 = \la3\,4\ra\,[\![3\,4]\!]$ can be replaced by combinations relevant for other channels by simultaneously replacing $\d^2\mu[\msf{s}]$ with $\d^2\mu[\msf{t}]$, or $\d^2\mu[\msf{u}]$. This simplifies \eqref{smhvym} to a single lightfront integral
\be\label{smhvym1}
\cA_\msf{s}^\text{MHV} = -\frac{2\,\rg^2}{\cN}\,\delta^3_{+,\perp}\!\left(\sum_{i=1}^4k_i\right)f^{\msf{a}_1\msf{a}_2\msf{b}}f^{\msf{a}_3\msf{a}_4\msf{b}}\frac{\la1\,2\ra^3}{\la3\,4\ra\la4\,1\ra}\frac{[\tilde\iota\,2]}{[\tilde\iota\,1]\la1\,3\ra}\int\d x^-\,\e^{\im\,\F_4}\,.
\ee
Similar IBP identities aid us in simplifying the other channels. For instance, the $\msf{u}$-channel contribution, initially found to be
\be\label{umhvym}
\begin{split}
\cA_\msf{u}^\text{MHV} &= 2\,\rg^2\,\delta^3\!\left(\sum_{r=1}^4k_r\right)f^{\msf{a}_1\msf{a}_4\msf{b}}f^{\msf{a}_2\msf{a}_3\msf{b}}\int\d^2\mu[\msf{u}]\;\frac{\la1\,2\ra[\tilde\iota\,4]}{\la3\,1\ra[\tilde\iota\,1][\tilde\iota\,2]}\,\biggl[\la1\,2\ra[\tilde\iota\,4][\![3\,2]\!](y^-)\\
&+ (\la1\,3\ra[\tilde\iota\,3]-\la1\,2\ra[\tilde\iota\,2])\left(\frac{\la\iota\,1\ra[\tilde\iota\,3][\![4\,1]\!](x^-)}{(k_1+k_4)_+} - \frac{\la\iota\,2\ra[\tilde\iota\,4][\![3\,2]\!](y^-)}{(k_2+k_3)_+}\right)\biggr]+ (x^-\leftrightarrow y^-)\,,
\end{split}
\ee
is simplified to
\be\label{umhvym1}
\cA_\msf{u}^\text{MHV}=-\frac{2\,\rg^2}{\cN}\,\delta^3_{+,\perp}\!\left(\sum_{i=1}^4k_i\right)f^{\msf{a}_1\msf{a}_4\msf{b}}f^{\msf{a}_2\msf{a}_3\msf{b}}\frac{\la1\,2\ra^3}{\la2\,3\ra\la4\,1\ra}\frac{[\tilde\iota\,4]}{[\tilde\iota\,1]\la1\,3\ra}\int\d x^-\,\e^{\im\F_4}\,,
\ee
using IBP identities and 3-momentum conservation. The $\msf{t}$-channel contribution 
\be\label{tmhvym}
\cA_\msf{t}^\text{MHV} = -\frac{2\,\rg^2}{\cN}\,\delta^3_{+,\perp}\!\left(\sum_{i=1}^4k_i\right)f^{\msf{a}_1\msf{a}_3\msf{b}}f^{\msf{a}_2\msf{a}_4\msf{b}}\frac{\la1\,2\ra^3}{\la2\,4\ra\la3\,1\ra}\frac{[\tilde\iota\,3]}{[\tilde\iota\,1]\la1\,4\ra}\int\d x^-\,\e^{\im\F_4}\,,
\ee
follows by exchanging particle labels $3$ and $4$. Note that it is highly non-trivial to obtain integrands to which these IBP identities can be applied; this is certainly not guaranteed by any na\"ive property of the Feynman diagrammatics.

The colour-ordered 4-gluon amplitude is now obtained by using the Jacobi identity $f^{\msf{a}_1\msf{a}_4\msf{b}}f^{\msf{a}_2\msf{a}_3\msf{b}} = -f^{\msf{a}_1\msf{a}_2\msf{b}}f^{\msf{a}_3\msf{a}_4\msf{b}} + f^{\msf{a}_1\msf{a}_3\msf{b}}f^{\msf{a}_2\msf{a}_4\msf{b}}$ to mix contributions. The result is
\be\label{fullmhvym}
\begin{split}
\cA_4^{\text{MHV}} &= \frac{2\,\rg^2}{\cN}\,\delta^3_{+,\perp}\!\left(\sum_{i=1}^4k_i\right)\int\d x^-\,\e^{\im\F_4}\\
&\times\left(f^{\msf{a}_1\msf{a}_2\msf{b}}f^{\msf{a}_3\msf{a}_4\msf{b}}\,\frac{\la1\,2\ra^4}{\la1\,2\ra\,\la2\,3\ra\,\la3\,4\ra\,\la4\,1\ra} + f^{\msf{a}_1\msf{a}_3\msf{b}}f^{\msf{a}_2\msf{a}_4\msf{b}}\,\frac{\la1\,2\ra^4}{\la1\,4\ra\,\la4\,2\ra\,\la2\,3\ra\,\la3\,1\ra}\right)\,,
\end{split}
\ee
which matches \eqref{MHVgluesdpw} for $n=4$ in the appropriate colour-ordering. 


\section{Impulsive SDPWs}
\label{App:Imp}

It is illustrative to consider our formula for the tree-level S-matrix in the limit where the SD radiative background gauge field becomes an \emph{impulsive} SDPW. This means that the background field strength has support only along a single lightfront hypersurface, representing an instantaneous burst of chiral colour radiation. This setting enables the $x_r$ moduli integrals of \eqref{YMamp-pw-fin} to be performed explicitly.

An impulsive SDPW located on the lightfront $x^-=0$ is described by a gauge field $A=-\tilde{\alpha}\,\Theta(x^-)\,\d\tilde{z}$, where $\tilde{\alpha}$ is a constant vector in a Cartan subalgebra of the gauge group and $\Theta(x^-)$ is the Heavyside step function. It is straightforward to show that the function $\cF(x_r)$ appearing in $g(x,\sigma)$ through \eqref{g-def2} is given by $\cF(x_r)=\tilde{\alpha}\,x_{r}\,\Theta(x_r)$ for each $r=1,\ldots,d$. Each $\d x_{r}$ integral in the formula \eqref{YMamp-pw-fin} for $\cA_{n,d}$ can now be performed by splitting 
\begin{equation*}
\int\limits_{-\infty}^{+\infty} \d x_{r}\,(\cdots)=\int\limits_{-\infty}^{0} \d x_{r}\,(\cdots)+\int\limits_{0}^{+\infty} \d x_{r}\,(\cdots)\,,
\end{equation*}
where an appropriate $\im\varepsilon$-prescription is assumed to ensure convergence of the integrals. These integrals can now be performed straightforwardly, with the result:
\begin{multline}\label{YMamp-imp}
\cA_{n,d}|_{\mathrm{impulsive}}=\delta\!\left(\sum_{i=1}^{n}[\tilde{o}\,i]\,\la\iota\,i\ra\right)\,\int\frac{\d^{d+1}b\,\d^{d+1}\nu}{\mathrm{vol}\,\mathrm{GL}(2,\C)}\,J(b,\nu)\,\delta^{1|4(d+1)}\!\left(\sum_{j=1}^{n}s_{j}\,([\tilde{\iota}\,j],\eta_j)\,\sigma_{j}^{\ba(d)}\right) \\
\times \mathcal{I}_{n,d}\, \prod_{i=1}^{n}\frac{\d s_i\,\D\sigma_i}{s_i\,(i\,i+1)}\,\bar{\delta}^{2}(\kappa_i-s_{i}\,\lambda(\sigma_i))\,,
\end{multline}
where the impulsive factor $\mathcal{I}_{n,d}$ is defined by:
\begin{multline}\label{Impfactor}
 \mathcal{I}_{n,d}:=\prod_{r=1}^{d}\lim_{\varepsilon\rightarrow 0}\left[\left(\nu_{r}\,\sum_{i=1}^{n}s_i\,[\tilde{o}\,i]\,\mathfrak{s}_{r}(\sigma_i) +\im\,\varepsilon\right)^{-1}\right. \\
\left. -\left(\nu_{r}\,\sum_{i=1}^{n}\left(s_i\,[\tilde{o}\,i]\,\mathfrak{s}_{r}(\sigma_i)-\im\,\frac{\tilde{\alpha}\,e_{i}}{b_0}\,\frac{(i\,\xi)}{(i\,b_r)\,(b_r\,\xi)}\prod_{s\neq r}\frac{1}{(b_r\,b_s)}\right)-\im\varepsilon\right)^{-1} \right]\,.
\end{multline}
In this expression, the role of the small parameter $\varepsilon$ is to regulate the flat background limit of the amplitudes.

Observe that in the weak field limit, where $\tilde{\alpha}\rightarrow 0$, the impulsive factor $\mathcal{I}_{n,d}$ obeys
\begin{equation}\label{weakIF}
\begin{split}
\lim_{\tilde{\alpha}\rightarrow 0}\mathcal{I}_{n,d} & = \prod_{r=1}^{d}\lim_{\varepsilon\rightarrow 0}\left[\frac{1}{\nu_{r}\,\sum_{i=1}^{n}s_i\,[\tilde{o}\,i]\,\mathfrak{s}_{r}(\sigma_i) +\im\,\varepsilon} -\frac{1}{\nu_{r}\,\sum_{i=1}^{n}s_i\,[\tilde{o}\,i]\,\mathfrak{s}_{r}(\sigma_i)-\im\varepsilon} \right] \\
 & = \prod_{r=1}^{d} \delta\!\left(\nu_{r}\sum_{i=1}^{n}s_i\,[\tilde{o}\,i]\,\mathfrak{s}_{r}(\sigma_i)\right)\,.
\end{split}
\end{equation}
Inserting this in \eqref{YMamp-imp} then reproduces the usual RSVW formula \eqref{RSVW} as expected, written in basis \eqref{ilamb1}, \eqref{olamb1} for the $\lambda_{\alpha}(\sigma)$ map components.


\section{Twistor action and MHV formalism}
\label{propapp}

The MHV formalism can be derived from the twistor action \eqref{YMTA1} description of perturbative Yang-Mills theory. For our purposes, we need a version of this action coupled to a SD background gauge field on space-time, which is given by shifting $a\mapsto\sa+a$,
\begin{multline}\label{YMTAapp}
S[a,b]=\frac{\im}{2\,\pi}\int_{\PT}\D^{3}Z\wedge\tr\left(b\wedge(\Dbar a+a\wedge a)\right)\\
 +\frac{\rg^2}{4}\!\int\limits_{\M\times\P^1\times\P^1}\!\!\!\d^{4}x\,\D\lambda_1\,\D\lambda_2\,\la \lambda_1\,\lambda_2\ra^2\,\tr\left[\widehat\sH^{-1}_1\,b_1\,\widehat\sH_{1}\,\widehat\sH^{-1}_{2}\,b_2\,\widehat\sH_{2}\right]\,,
\end{multline}
where $\Dbar=\dbar+\sa$ gives an integrable background complex structure on the Ward bundle $E\to\PT$, and $\widehat\sH$ solves \eqref{pertframedef}. The MHV vertices can be computed by expanding the second term of \eqref{YMTAapp} as a series in $a$ around the background $\sa$. This calculation was already done from a generating functional viewpoint to arrive at \eqref{TYMgen3}. Thus, to furnish MHV rules, we are left to derive the MHV propagator. In this appendix, we illustrate its derivation for a SDPW background (cf., \cite{Jiang:2008xw,Adamo:2013cra} for the derivation in a trivial background).

The following analysis is best carried out in Euclidean signature on space-time, $\R^4$, so that the twistor space is identified with its projective spinor bundle $\PT\simeq\R^4\times\P^1$. In coordinates $(x,\lambda)$ on this space, one can re-express the complex structure of twistor space as~\cite{Woodhouse:1985id}
\be\label{dbardef}
\dbar = \bar e^0\,\dbar_0 + \bar e^{\dal}\,\dbar_{\dal}\,,
\ee
where, introducing quaternionic conjugation $\hat\lambda_\alpha = (\overline{\lambda_1},-\overline{\lambda_0})$, one defines a standard basis of $(0,1)$-forms,
\be\label{e0eal}
\bar e^0 = \frac{\D\hat\lambda}{\la\lambda\,\hat\lambda\ra^2} = \frac{\la\hat\lambda\,\d\hat\lambda\ra}{\la\lambda\,\hat\lambda\ra^2}\,,\qquad\bar e^{\dal} = \frac{\hat\lambda_\al\,\d x^{\al\dal}}{\la\lambda\,\hat\lambda\ra}\,,
\ee
as well as a basis of $(0,1)$-vector fields,
\be\label{d0dal}
\dbar_0 = \la\lambda\,\hat\lambda\ra\,\lambda^\al\frac{\p}{\p\hat\lambda^\al}\,,\qquad\dbar_{\dal} = \lambda^\al\p_{\al\dal}\,.
\ee
The Ward bundle also gets identified with the pullback of the Yang-Mills bundle by the natural projection $\PT\to\R^4$. The gauge connection $A_a$ on $\R^4$ lifts to a partial connection on $E\to\PT$,\footnote{This representative for the partial connection $\sa$ is gauge equivalent on twistor space to the one given by \eqref{radiativeym}.} 
\be\label{awoodhouse}
\sa = -\im\,\lambda^\al A_{\al\dal}\,\bar e^{\dal}\,.
\ee
This defines our background coupled complex structure $\Dbar = \bar e^0\,\dbar_0 + \bar e^{\dal}\,\lambda^\al D_{\al\dal}$, where $D_a = \p_a - \im\,A_a$ is the space-time covariant derivative.

To compute the propagator, consider the kinetic terms in \eqref{YMTAapp} and add to these source terms for $a$ and $b$,
\be\label{absource}
\frac{\im}{2\,\pi}\int_{\PT}\D^{3}Z\wedge\tr\left(b\wedge\Dbar a + a\wedge J + b\wedge K\right)\,,
\ee
with $(0,2)$-form sources $J$ and $K$ of weights $-4$ and $0$ in $\lambda$ respectively. This produces the background coupled linear field equations,
\be
\Dbar a = -K\,,\qquad\Dbar b = -J\,.
\ee
Next, expand
\be
a = a_0\,\bar e^0 + a_{\dal}\,\bar e^{\dal}\,,\qquad K = \frac{1}{2}\,K_0\,\bar e^{\dal}\wedge\bar e_{\dal} - K_{\dal}\,\bar e^0\wedge\bar e^{\dal}\,,
\ee
so that the first of these equations becomes
\be
\lambda^\al D_{\al}^{\dal}a_{\dal} = K_0\,,\qquad\dbar_0 a_{\dal} - \lambda^\al D_{\al\dal}a_0 = -K_{\dal}\,.
\ee
Choosing the most natural CSW gauge $\tilde\iota^{\dal}\p_{\dal}\lrcorner a  = 0$ and transforming to momentum space,
\be
a(x,\lambda) = \int\frac{\d^4k}{(2\pi)^4}\;\hat a(k,\lambda)\,\e^{\im\phi(x)}\,,\qquad K(x,\lambda) = \int\frac{\d^4k}{(2\pi)^4}\;\hat K(k,\lambda)\,\e^{\im\phi(x)}\,,
\ee
these are straightforwardly inverted to yield
\be\label{aosrep}
\hat a= -\frac{4\pi\im}{k^2}\,\bar\delta^1_{-2}(\la\lambda|k|\tilde\iota])\,\hat K_0 - \frac{\im\,\tilde\iota^{\dal}\,\hat K_{\dal}\,\bar e^0}{\la\lambda|k|\tilde\iota]} + \frac{\im\,\hat K_0\,\tilde\iota_{\dal}\,\bar e^{\dal}}{\la\lambda|k|\tilde\iota]}\,.
\ee
Here, $\phi(x)$ is the off-shell continuation of the phase given in \eqref{sdpwHJ}, and the weighted delta functions $\bar\delta^1_m(\la\lambda\,\kappa\ra)$ are defined in \eqref{delta1m}. To facilitate the computation, one can define an off-shell dressed momentum again by $K_{\al\dal} = \p_{\al\dal}\phi - e\,A_{\al\dal}$ which obeys $K^2=k^2$ and $\tilde\iota^{\dal} K_{\al\dal} = \tilde\iota^{\dal}k_{\al\dal}$. 

A similar solution can be found for $b$ in momentum space,
\be\label{bosrep}
\hat b= -\frac{4\pi\im}{k^2}\,\bar\delta^1_{-2}(\la\lambda|k|\tilde\iota])\,\hat J_0 - \frac{\im\,\tilde\iota^{\dal}\,\hat J_{\dal}\,\bar e^0}{\la\lambda|k|\tilde\iota]} + \frac{\im\,\hat J_0\,\tilde\iota_{\dal}\,\bar e^{\dal}}{\la\lambda|k|\tilde\iota]}\,,
\ee
having accounted for homogeneities. With \eqref{aosrep}, \eqref{bosrep} it is straightforward to construct the MHV propagator as a Green's function for $\Dbar$, giving \eqref{mhvprop}.

\bibliography{sdpw1}
\bibliographystyle{JHEP}

\end{document}